\newif\ifpreprint
 
\preprinttrue 
  
\ifpreprint
\documentclass[journal=jctcce,manuscript=article]{achemso}
\else
\documentclass[journal=jctcce,manuscript=article,layout=twocolumn]{achemso}
\fi

\setkeys{acs}{articletitle = false}

\usepackage{amsmath}
\usepackage{amssymb}
\usepackage{mathtools}
\usepackage{commath}

\usepackage{newtxtext,newtxmath}
\usepackage{empheq}

\usepackage{braket}
\usepackage[normalem]{ulem} 
\usepackage{threeparttable}
\usepackage[version=4]{mhchem} 
\usepackage{indentfirst}
\usepackage{color,soul}
\usepackage{cases}
\usepackage{array} 

\usepackage{xcolor}
\usepackage{xspace}
\usepackage{ifthen}

\usepackage[colorlinks = true,
            linkcolor = blue,
            urlcolor  = black,
            citecolor = blue,
            anchorcolor = black]{hyperref}

\definecolor{goodorange}{RGB}{225,125,0}
\definecolor{goodgreen}{RGB}{5,130,5}
\definecolor{goodred}{RGB}{220,50,25}
\definecolor{goodblue}{RGB}{30,144,255}
\definecolor{OliveGreen}{RGB}{5,100,5}

\newcommand*{\kcal}{kcal mol$^{-1}$\xspace}
\newcommand*{\sunit}{$E_{\rm h}^{-2}$\xspace}
\newcommand*{\molpro}{{\scshape Molpro}\xspace}

\newcommand*{\PSI}{{\scshape Psi4}\xspace}

\newcommand*{\forte}{{\scshape Forte}\xspace}

\newcommand{\mref}[0]{\Psi_0}
\newcommand{\tens}[3]{{#1}_{#2}^{#3}}
\newcommand{\dfock}[1]{\epsilon_{#1}}
\newcommand{\cop}[1]{\hat{a}^\dag_{#1}}
\newcommand{\aop}[1]{\hat{a}_{#1}}
\newcommand{\sqop}[2]{\hat{a}_{#2}^{#1}}

\newcommand{\aphystei}[2]{\bra{#1}\!\!\ket{#2}}
\newcommand{\kro}[2]{\delta_{#2}^{#1}}
\newcommand{\mpd}[2]{\Delta_{#2}^{#1}}
\newcommand{\density}[2]{\gamma_{#2}^{#1}}

\newcommand{\mdensity}[2]{\tilde{\gamma}_{#2}^{#1}}
\newcommand{\edens}[2]{\Gamma_{#2}^{#1}}
\newcommand{\cdensity}[2]{\eta_{#2}^{#1}}
\newcommand{\cumulant}[2]{\lambda_{#2}^{#1}}

\newcommand{\orbrot}[2]{\vartheta_{#2}^{#1}}
\newcommand{\no}[1]{ \{ {#1} \}}

\newcommand{\rh}[0]{\tilde{h}}

\newcommand{\note}[2]{
\ifthenelse{\equal{#1}{S}}{
\colorbox{goodblue}{\textcolor{white}{\footnotesize \fontfamily{phv}\selectfont #1}}
    \textcolor{goodblue}{{\footnotesize \fontfamily{phv}\selectfont #2}}\xspace
}{}
\ifthenelse{\equal{#1}{F}}{
\colorbox{goodorange}{\textcolor{white}{\footnotesize \fontfamily{phv}\selectfont #1}}
    \textcolor{goodorange}{{\footnotesize \fontfamily{phv}\selectfont #2}}\xspace
}{}
\ifthenelse{\equal{#1}{Y}}{
\colorbox{goodred}{\textcolor{white}{\footnotesize \fontfamily{phv}\selectfont #1}}
    \textcolor{goodred}{{\footnotesize \fontfamily{phv}\selectfont #2}}\xspace
}{}
}

\usepackage{titlesec}

\usepackage[fontsize=11pt]{scrextend}
\titleformat{\section}
{\normalfont\large\sffamily\bfseries}
{\thesection.}{0.5em}{\uppercase}

\titleformat{\subsection}[runin]
{\normalfont\sffamily\bfseries}
{\thesubsection}{0.25em}{}[.\;\;]

\titleformat{\subsubsection}[runin]
{\normalfont\sffamily\itshape}
{\thesubsubsection}{0.25em}{}[.\;\;]

\titleformat{\suppinfo}
{\normalfont\sffamily\bfseries}
{\thesubsection}{0.25em}{}

\titlespacing*{\section}{0pt}{0.5\baselineskip}{0.1\baselineskip}
\titlespacing*{\subsection}{0pt}{0.25\baselineskip}{0.1\baselineskip}
\titlespacing*{\subsubsection}{0pt}{0.1\baselineskip}{0.0\baselineskip}

\NewDocumentCommand{\dev}{sO{}mm}{
  \IfBooleanTF{#1}
    {{\rm d^{#2}} {#3} / {\rm d} {#4}^{#2}}
    {\frac{{\rm d^{#2}} {#3}}{{\rm d} {#4}^{#2}}}
}

\NewDocumentCommand{\pdev}{sO{}mm}{
  \IfBooleanTF{#1}
    {\partial^{#2} {#3} / \partial {#4}^{#2}}
    {\frac{\partial^{#2} {#3}}{\partial {#4}^{#2}}}
}

\newcommand{\exps}[2]{e^{-s(\Delta_{#1}^{#2})^2}}
\newcommand{\indicator}[2]{{\delta}_{{#2} \in \mathbb{#1}}}

\author{Shuhe Wang}
\email{shuhe.wang@emory.edu}
\affiliation{Department of Chemistry and Cherry Emerson Center for Scientific Computation, Emory University, Atlanta, GA 30322, USA}

\author{Chenyang Li}
\email{chenyang.li@bnu.edu.cn}
\affiliation{Key Laboratory of Theoretical and Computational Photochemistry, Ministry of Education, College of Chemistry, Beijing Normal University, Beijing 100875, China}

\author{Francesco A. Evangelista}
\email{francesco.evangelista@emory.edu}
\affiliation{Department of Chemistry and Cherry Emerson Center for Scientific Computation, Emory University, Atlanta, GA 30322, USA}

\let\oldmaketitle\maketitle
\let\maketitle\relax

\title{Analytic Energy Gradients for the Driven Similarity Renormalization Group Multireference Second-Order Perturbation Theory}

\begin{document}

\ifpreprint
\else
\twocolumn[
\begin{@twocolumnfalse}
\fi
\oldmaketitle

\begin{abstract}
We derive analytic energy gradients of the driven similarity renormalization group (DSRG) multireference second-order perturbation theory (MRPT2) using the method of Lagrange multipliers.
In the Lagrangian, we impose constraints for a complete-active-space self-consistent-field reference wave function and the semicanonical orthonormal molecular orbitals.
Solving the associated Lagrange multipliers is found to share the same asymptotic scaling of a single DSRG-MRPT2 energy computation.
A pilot implementation of the DSRG-MRPT2 analytic gradients is used to optimize the geometry of the singlet and triplet states of {\it p}-benzyne.
The equilibrium bond lengths and angles are similar to those computed via other MRPT2s and Mukherjee's multireference coupled cluster theory.
An approximate DSRG-MRPT2 method that neglects the contributions of three-body density cumulant is found to introduce negligible errors in the geometry of {\it p}-benzyne, lending itself to a promising low-cost approach for molecular geometry optimizations using large active spaces.
\end{abstract}

\ifpreprint
\else
\end{@twocolumnfalse}
]
\fi

\ifpreprint
\else
\small
\fi

\noindent

\section{Introduction}
\label{sec:intro}

Analytic energy derivatives play a central role in modern quantum chemistry.\cite{abbott2021arbitrary}
They enable efficient geometry optimizations and \textit{ab initio} (including non-adiabatic) molecular dynamics simulations,\cite{Iftimie.2005,Park:2017jb,Curchod.2018} two tasks that require rapid evaluation of energy gradients with respect to nuclear coordinates.
Recent developments of analytic gradients for local correlation methods have extended first principles geometry optimizations to weakly correlated molecules with hundreds of nuclear degrees of freedom.\cite{Pinski2018,Dornbach2019,Ni:2019cd}
For strongly correlated systems (e.g., diradicals and transition-metal complexes), multireference (MR) methods\cite{Szalay:2012df,Lyakh:2012cn,Koehn:2013cp,Evangelista:2018bt} are generally necessary to obtain accurate global potential energy surfaces (PESs).
Unfortunately, the development of analytic energy gradients for multireference theories has trailed that of single-reference methods both in terms of the underlying mathematical formalism and the broad availability of software implementations, limiting studies of strongly correlated systems.

Within the domain of MR methods, the sweet spot between accuracy and computational cost is found in second-order perturbation theory (MRPT2).
Various MRPT2 methods have been proposed over the years,\cite{andersson1990second,hirao1992multireference,andersson1992second,Kozlowski:1994cw,werner1996third,Mahapatra:1999gh,angeli2001introduction,*angeli2002n,Khait:2002ca,Szabados:2005jk,chaudhuri2005comparison,hoffmann2009comparative,evangelista2009companion,Sokolov:2017gr,giner2017jeziorski}
among which the most widely applied are the complete-active-space (CAS) second-order perturbation theory (CASPT2)\cite{andersson1992second} and \textit{n}-electron valence second-order perturbation theory (NEVPT2).\cite{angeli2001introduction,*angeli2002n}
The CASPT2 scheme based on a single CAS configuration interaction (CI) state is known to suffering from the intruder-state problem.
This issue is commonly addressed by applying level shifts to the diagonal elements of the one-body zeroth-order Hamiltonian.\cite{roos1995multiconfigurational,Forsberg:1997ke}
A different, parameter-free approach is used in NEVPT2 to deal with intruder states, whereby the zeroth-order Hamiltonian is augmented with bi-electronic terms, as proposed by Dyall.\cite{Dyall:1995ct}
Nonetheless, both CASPT2 and NEVPT2 in principle require the four-body reduced density matrix (4-RDM) of the CASCI wave function that are both costly to compute and store in memory.
Numerous efforts have been made to reduce the cost of high-order density matrices.
For example, building and storing the 4-RDM can be avoided using a cumulant decomposition\cite{Valdemoro.1992,Colmenero.1993,kutzelnigg1997normal,Mazziotti.1998,Mazziotti:2006gf,Shamasundar:2009ee,Misiewicz:2020ia} and subsequently neglecting contributions from the 4-body density cumulant.
This approach has lent itself to efficient and robust implementations of CASPT2 that can handle up to thirty active orbitals.\cite{Kurashige:2014bq,Phung:2016ke}
Such approximations are less successful in NEVPT2 and ``false intruders'' may appear due to the density dependencies of the Koopman's matrices in the energy denominators.\cite{Zgid:2009fu,Guo2021a,*Guo2021b}
For a similar reason, the use of Cholesky decomposed integrals may also destabilize the numerical robustness of the NEVPT2 method.\cite{Freitag:2017ir}
For certain formulations of MRPT2, it is possible to avoid computing the 4-RDM by introducing appropriate intermediates, employing an uncontracted formalism, or a matrix product state reference.\cite{sharma2014communication,sharma2017combining,sokolov2016time,Sokolov:2017gr,sokolov2018multi,chatterjee2020extended}

The developments of analytic energy gradients for MRPT2s were largely overlooked for a long time.
The very first derivation were reported by Nakano and co-workers in 1998,\cite{nakano1998analytic} on the multi-configurational quasi-degenerate perturbation theory (MC-QDPT).\cite{Nakano:1993hv}
However, applications of MC-QDPT gradient theory were restricted to small systems\cite{Nakano:2011gl} until the recent work of Park that employs the analytic gradient theory of the extended MC-QDPT to optimize the conical intersections of a retinal model chromophore.\cite{Park2021}
The analytic first derivatives have also been developed for Werner's partially contracted CASPT2\cite{celani2003analytical} and Hoffmann's generalized Van Vleck perturbation theory,\cite{Dudley2003,Theis:2011bd} along with their extensions for excited states.\cite{Khait:2012fi,shiozaki2011communication,Gyorffy:2013kf}

More recently, significant advances have been made in developing the analytic gradients for CASPT2 and NEVPT2.
The analytic gradients for the fully internally contracted CASPT2 were first achieved by MacLeod and Shiozaki via automatic code generation.\cite{macleod2015communication}
Multi-state generalizations of CASPT2 have also been derived by Shiozaki and co-workers,\cite{Vlaisavljevich:2016kv,Park:2017je,Park:2019fh} and made publicly available through the BAGEL package.\cite{Shiozaki:2018fm}
Song, Mart{\'i}nez, and Neaton developed the analytic gradients for the reduced scaling CASPT2 based on supporting subspace method.\cite{Song:2021hp,Song2021}
Gradient theory for NEVPT2 was introduced independently by Park\cite{Park2019,Park2020} and Nishimoto.\cite{Nishimoto:2019hd}

The driven similarity renormalization group (DSRG) provides an alternative framework to formulate MR theories that avoid the intruder-state problem and yield smooth PESs.\cite{evangelista2014driven,Li:2019fu}
In the DSRG, the many-body Hamiltonian is unitarily transformed in such a way that interactions that couple the reference state and the excited configurations are zeroed (this is equivalent to a unitary internally contracted theory).
Importantly, this decoupling depends on the magnitude of the energy denominator of each interaction removed, and it is gradually suppressed when a denominator approaches zero.
This feature of the DSRG introduces a separation of energy scales, the extent of which is controlled via the so-called flow parameter $s$.
For finite values of $s$, the MR-DSRG methods yield continuous potential energy surfaces that are free from the characteristic ``spikes'' caused by intruder states.

Over the past few years, we have proposed and implemented several practical MR-DSRG ans{\"a}tze.\cite{li2015multireference,Li:2016hb,li2017driven,Zhang:2019ec}
The least computational demanding member of this family is the DSRG-MRPT2 method.\cite{li2015multireference}
This approach uses a diagonal normal-ordered Fock operator as the zeroth-order Hamiltonian.
As a result, the DSRG-MRPT2 energy only depends on the reference 1-, 2-, and 3-RDMs.
Previous benchmarks on small molecules show that the DSRG-MRPT2 approach yields PESs of similar accuracy to other MRPT2s.\cite{li2017driven,Li:2021a}
When combined with factorization of the two-electron integrals, DSRG-MRPT2 be routinely applied to systems with more than two thousand basis functions.\cite{hannon2016integral}
The DSRG-MRPT2 approach has also been combined with approximate CASCI methods to target large active spaces.\cite{Schriber:2018hw,Khokhlov2021}
These encouraging results motivate us to further extend its applicability.

Herein, we report a pilot implementation of the analytic energy gradients for the state-specific unrelaxed DSRG-MRPT2 method.\cite{li2015multireference}
The DSRG-MRPT2 energy is not variationally optimized with respect to the orbital and CI coefficients, nor the cluster amplitudes.
Following a standard approach,\cite{helgaker1988analytical,Helgaker:2002hs} we construct a Lagrangian function ($\cal L$) and incorporate constraints for non-variational quantities.
As anticipated from previous experiences,\cite{celani2003analytical,macleod2015communication,wang2019analytic} 
the computational bottleneck of the gradient procedure is solving the coupled Z-vector equations\cite{handy1984evaluation,Yamaguchi1994book} for the orbital and CI coefficients.
Due to the complexity of the DSRG-MRPT2 analytic gradients, in this work we restrict our derivation to the original unrelaxed approach.\cite{li2015multireference}
Variants of the DSRG-MRPT2 that include reference relaxation\cite{li2017driven} will be considered in future works.

This paper is organized as follows.
We start by introducing the DSRG-MRPT2 energy expressions and the amplitude equations in Sec.~\ref{subsec:energy}, followed by a general discussion of gradient theory using the method of Lagrange multipliers in Sec.~\ref{subsec:gradient}.
We report expressions for all the constraints and the corresponding Lagrange multipliers in Sec.~\ref{subsec:constraints} and \ref{subsec:multipliers}, respectively.
The theory section is concluded with a brief discussion on the computational cost and limitations of the current implementation (see Sec.~\ref{subsec:impl}).
In Sec.~\ref{sec:results}, we report the adiabatic singlet--triplet splittings of \textit{p}-benzyne computed from the DSRG-MRPT2 optimized geometries using analytic gradients.
Finally, we conclude this work in Sec.~\ref{sec:conclusions} by pointing out several future directions and applications of DSRG-MRPT2 gradient theory.

\section{Theory}
\label{sec:theory}

We first introduce the orbital notation adopted in this work.
Consider a set of CASSCF orthonormal molecular spin orbitals (MSOs) $\mathbb{G} \equiv \{\psi_p ({\bf r}, \omega) = \phi_p ({\bf r}) \sigma_p(\omega), p = 1, 2, \dots, N_{\rm G} \}$.
Each MSO is a product of a molecular orbital (MO) $\phi_p (\bf r)$ and a spin function $\sigma_p(\omega)$,
and the spatial and spin coordinates are indicated with $\bf r$ and $\omega$, respectively.
An MO is a linear combination of nonorthogonal atomic orbitals (AOs) $\chi_\mu ({\bf r})$:
\begin{equation}
\label{eq:mo}
\phi_p({\bf r}) = \sum_{\mu}^{\rm AO}  \chi_\mu({\bf r}) \, C_{\mu p},
\end{equation}
where $C_{\mu p}$ is the orbital coefficient matrix.
The MSOs are assumed to be orthonormal, in which case the MSO overlap integral ($\tens{S}{p}{q}$) is an identity matrix:
\begin{equation}
\label{eq:mo_overlap}
\tens{S}{p}{q} = \braket{\psi_p | \psi_q} = \kro{q}{p},
\end{equation}
where $\kro{q}{p}$ is the Kronecker delta.
We partition the MSOs into three subsets: core ($\mathbb{C}$, doubly occupied), active ($\mathbb{A}$, partially occupied), and virtual ($\mathbb{V}$, unoccupied).
For convenience, we also introduce composite orbital spaces, namely, hole ($\mathbb{H} = \mathbb{C} \cup \mathbb{A}$) and particle ($\mathbb{P} = \mathbb{A} \cup \mathbb{V}$).
The indices labeling MSOs are summarized in Table \ref{tab:orbital_space}, and Greek letters $\mu, \nu, \rho, \tau$ are utilized to index AOs.

\begin{table}[ht!]
\caption{Partition of the spin orbital spaces.}
\label{tab:orbital_space}
\ifpreprint
\renewcommand{\arraystretch}{1.0}
\begin{tabular*}{0.75\columnwidth}{@{\extracolsep{\stretch{1}}} l c c c c @{}}
\else
\renewcommand{\arraystretch}{1.1}
\footnotesize
\begin{tabular*}{\columnwidth}{@{\extracolsep{\stretch{1}}} l c c c c @{}}
\fi
\hline
\hline
Space & Symbol & Size & Indices & Description \\
\hline
Core & $\mathbb{C}$ & $N_{\rm C}$ & $m,n,o$ & Occupied \\
Active & $\mathbb{A}$ & $N_{\rm A}$ & $u,v,w,x,y,z$ & Partially occupied \\
Virtual & $\mathbb{V}$ & $N_{\rm V}$ & $e,f$ & Unoccupied \\
Hole & $\mathbb{H}$ & $N_{\rm H}$ & $i,j,k,l$ & {$\mathbb{C} \cup \mathbb{A}$} \\
Particle & $\mathbb{P}$ & $N_{\rm P}$ & $a,b,c,d$ & {$\mathbb{A} \cup \mathbb{V}$}\\
General & $\mathbb{G}$ & $N_{\rm G}$ & $p, q, r, s$ & {$\mathbb{C} \cup \mathbb{A} \cup \mathbb{V}$} \\
\hline
\hline
\end{tabular*}
\end{table}

The CASSCF reference wave function $\mref$ (often referred to as the ``reference'' in the following) is a linear combination of Slater determinants $\Phi_I$:
\begin{equation}
\label{eq:cas_wfn}
\ket{\mref} = \sum_{I}^{{\cal M}_0} c_I \ket{\Phi_I},
\end{equation}
with $c_I$ being the vector of CI coefficients.
These determinants form a complete active space (CAS) denoted by ${\cal M}_0$.
Any $\Phi_I \in {\cal M}_0$ can be expressed as
\begin{equation}
\label{eq:det}
\ket{\Phi_I} = \hat{\cal I}^\dagger \prod_{m}^{\mathbb{C}} \cop{m} \ket{-},
\end{equation}
where $\ket{-}$ is the true vacuum and $\cop{p}$ ($\aop{p}$) is a fermionic creation (annihilation) operator.
In Eq.~\eqref{eq:det}, the operator $\hat{\cal I}^\dagger = \cop{u} \cop{v} \cdots$ is one of the $|{\cal M}_0|$ choices of creating $n_a$ electrons in the $N_{\rm A}$ active orbitals in such a way that $\Phi_I$ has the desired spin and spatial symmetry.
In the following, we use capital letters $I$ and $J$ to label the index of determinants in ${\cal M}_0$.

It is convenient to express the properties of the reference in terms of general $n$-particle reduced density matrices ($n$-pRDMs), with elements defined as 
\begin{equation}
\label{eq:rdm}
\density{kl\cdots}{ij\cdots} = \braket{\mref| \underbrace{\cop{k} \cop{l} \cdots}_{n\text{ operators}} \, \underbrace{\cdots \aop{j} \aop{i}}_{n\text{ operators}} |\mref}.
\end{equation}
For example, the reference energy $E_0 = \braket{\mref| \hat{H} |\mref}$ may be expressed in terms of the 1- and 2-pRDMs ($\density{u}{v}$ and $\density{uv}{xy}$), and the one-electron ($\tens{h}{p}{q}$) and antisymmetrized two-electron ($\tens{v}{pq}{rs}$) integrals.
In particular, we have
\begin{equation}
\label{eq:E0ref}
E_0 = \braket{\mref| \hat{H} |\mref} =  E_0^{\rm c} + E_0^{\rm a},
\end{equation}
where the core ($E_0^{\rm c} $) and active ($E_0^{\rm a}$) parts of the energy are defined as:
\begin{align}
E_0^{\rm c} &= \sum_{m}^{\mathbb{C}} \tens{h}{m}{m} + \frac{1}{2} \sum_{mn}^{\mathbb{C}} \tens{v}{mn}{mn}, \label{eq:E0c} \\
E_0^{\rm a} &= \sum_{uv}^{\mathbb{A}} \tens{\bar{f}}{u}{v} \density{u}{v} + \frac{1}{4} \sum_{uvxy}^{\mathbb{A}} \tens{v}{uv}{xy} \density{uv}{xy}. \label{eq:E0a}
\end{align}
In Eq.~\eqref{eq:E0a}, we have introduced the core Fock matrix ($\tens{\bar{f}}{p}{q}$):
\begin{equation}
\label{eq:Fc}
\tens{\bar{f}}{p}{q} = \tens{h}{p}{q} + \sum_m^{\mathbb{C}} \tens{v}{pm}{qm}.
\end{equation}

\begin{figure}[ht!]
\centering
\ifpreprint
\includegraphics[width=0.25\columnwidth]{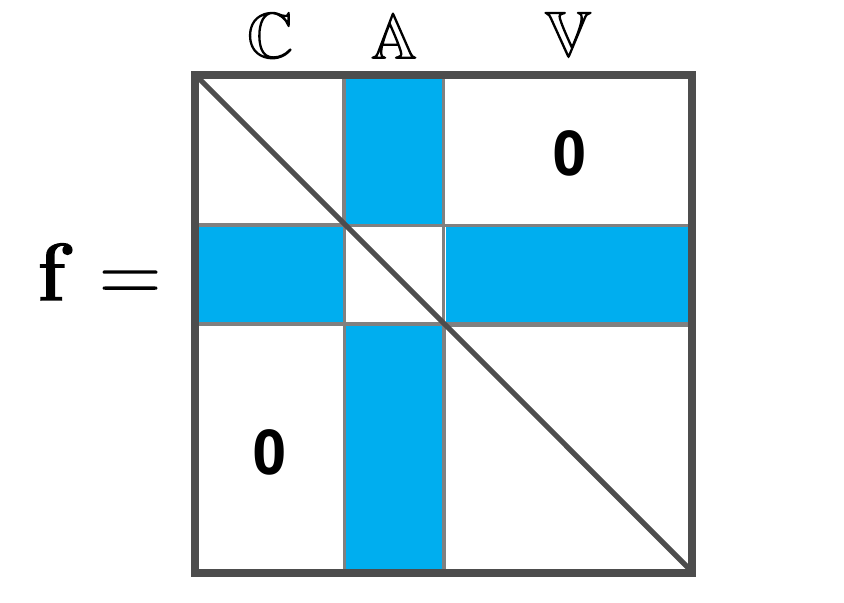}
\else
\includegraphics[width=0.5\columnwidth]{figure1}
\fi
\caption{Generalized Fock matrix in the semicanonical CASSCF basis. 
The blocks colored in blue are dense while the three diagonal blocks contain only diagonal elements.}
\label{fig:fockmat}
\end{figure}

We further define a generalized Fock matrix of the reference with components $\tens{f}{q}{p}$ given by:
\begin{equation}
\label{eq:fock}
\tens{f}{q}{p} = \tens{\bar{f}}{q}{p} + \sum_{uv}^{\mathbb{A}} \tens{v}{qv}{pu} \density{v}{u}.
\end{equation}
The DSRG-MRPT2 method is formulated in the semicanonical orbital basis, such that the core, active, and virtual blocks of the generalized Fock matrix (see Fig.~\ref{fig:fockmat}) are diagonal:
\begin{equation}
\label{eq:semi}
\tens{f}{p}{q} = \tens{f}{p}{p} \kro{q}{p}, \quad \forall\, p, q \in \mathbb{O}, \,\, \forall\, \mathbb{O} \in \{\mathbb{C} , \mathbb{A}, \mathbb{V}\}.
\end{equation}
The diagonal entries $\tens{f}{p}{p}$ can thus be viewed as orbital energies and they are denoted as $\dfock{p}$.
From here, the M{\o}ller--Plesset denominators are defined as:
\begin{equation}
\label{eq:mp_denorm}
\mpd{ij\cdots}{ab\cdots} = \dfock{i} + \dfock{j} + \cdots - \dfock{a} - \dfock{b} - \cdots.
\end{equation}
In this semicanonical basis, the zeroth-order Hamiltonian [$\hat{H}^{(0)}$] of DSRG-MRPT2 has a simple form:
\begin{equation}
\hat{H}^{(0)} = E_0 + \sum_{p}^{\mathbb{G}} \dfock{p} \no{ \cop{p} \aop{p} },
\end{equation}
where the curly braces ``$\{\cdot\}$'' indicate operator normal ordering with respect to the state $\mref$, following the approach of Mukherjee and Kutzelnigg.\cite{kutzelnigg1997normal}

\subsection{DSRG-MRPT2 Energy}
\label{subsec:energy}

In this section, we summarize the DSRG-MRPT2 energy expression within the unrelaxed formalism.
The reader is encouraged to consult Ref.~\citenum{li2015multireference} for a detailed derivation.
In Table \ref{tab:notation_energy}, we summarize the scalar and tensorial quantities that enter in the DSRG-MRPT2 energy expression.

\begin{table}[ht!]
\caption{Summary of notations used in the DSRG-MRPT2 energy.}
\label{tab:notation_energy}
\ifpreprint
\renewcommand{\arraystretch}{1.1}
\begin{tabular*}{0.75\columnwidth}{@{\extracolsep{\stretch{1}}} l l l @{}}
\else
\renewcommand{\arraystretch}{1.35}
\footnotesize
\begin{tabular*}{\columnwidth}{@{\extracolsep{\stretch{1}}} l l l @{}}
\fi
\hline
\hline
Name & Expression & Description \\
\hline
$E_0$ & Eq.~\eqref{eq:E0ref} & CASCI reference energy \\
$E^{(2)}$ & Eq.~\eqref{eq:E2nd} & DSRG second-order energy correction \\
\hline
$\tens{h}{p}{q}$ & $\braket{\psi_p|\hat{h}|\psi_q}$ & 1-electron integrals \\
$\tens{v}{pq}{rs}$ & $\aphystei{\psi_p \psi_q}{\psi_r \psi_s}$ & antisymmetrized 2-electron integrals \\
$\tens{\bar{f}}{p}{q}$ & Eq.~\eqref{eq:Fc} & core Fock matrix \\
$\tens{f}{p}{q}$ & Eq.~\eqref{eq:fock} & generalized Fock matrix \\
$\tens{\check{f}}{i}{a}$ & Eq.~\eqref{eq:fock_check} & modified first-order Fock matrix \\
\hline
$\density{kl\cdots}{ij\cdots}$ & Eq.~\eqref{eq:rdm} & $n$-particle reduced density matrices \\
$\mpd{ij\cdots}{ab\cdots}$ & Eq.~\eqref{eq:mp_denorm} & M{\o}ller--Plesset denominators \\
$\tens{\tilde{h}}{ij\cdots}{ab\cdots}$ & Eqs.~\eqref{eq:ren_f} \& \eqref{eq:ren_v} & modified first-order integrals \\
$\tens{t}{ab\cdots}{ij\cdots}$ & Eqs.~\eqref{eq:amp_1} \& \eqref{eq:amp_2} & first-order cluster amplitudes \\
\hline
\hline
\end{tabular*}
\end{table}

The unrelaxed DSRG-MRPT2 energy ${\cal E}(s)$ is the sum of the the reference energy $E_0$ [Eq.~\eqref{eq:E0ref}] and a second-order correction $E^{(2)}(s)$:
\begin{equation}
\label{eq:energy_short}
{\cal E}(s) = E_0 + E^{(2)} (s),
\end{equation}
where $s \in [0, +\infty)$ is the DSRG flow parameter, whose significance will be clarified later.
The second-order energy correction in Eq.~\eqref{eq:energy_short} is given by the fully contracted terms from an effective first-order Hamiltonian $\widetilde{H}^{(1)} (s)$ and a first-order cluster operator $\hat{T}^{(1)} (s)$:
\begin{equation}
\label{eq:E2nd}
E^{(2)} (s) = \braket{\mref| [\widetilde{H}^{(1)} (s), \hat{T}^{(1)} (s)] |\mref}.
\end{equation}
Detailed expressions for equation~\eqref{eq:E2nd} are presented in Appendix \ref{app:energy}.
In general, $E^{(2)} (s)$ is a sum of tensor contractions of the first-order cluster amplitudes [$\tens{t}{a}{i,(1)}(s)$, $\tens{t}{ab}{ij,(1)}(s)$],
the modified first-order integrals [$\tens{\rh}{i}{a,(1)}(s)$, $\tens{\rh}{ij}{ab,(1)}(s)$],
and 1-, 2- and 3-pRDMs.
For brevity, in the following we drop the superscript ``(1)'' for the first-order quantities and the label ``($s$)'' for $s$-dependent amplitudes or integrals.

The effective first-order Hamiltonian possesses the form
\begin{equation}
\label{eq:widetildeH}
\widetilde{H} = \sum_{i}^{\mathbb{H}} \sum_{a}^{\mathbb{P}} \tens{\tilde{h}}{i}{a} \no{ \sqop{i}{a} }
+ \frac{1}{4} \sum_{ij}^{\mathbb{H}} \sum_{ab}^{\mathbb{P}} \tens{\tilde{h}}{ij}{ab} \no{ \sqop{ij}{ab} },
\end{equation}
where the modified first-order integrals are given by:\cite{li2015multireference}
\begin{align}
\tens{\rh}{i}{a} &= \tens{f}{i}{a} + \tens{\check{f}}{i}{a} - \mpd{i}{a} \tens{t}{a}{i}, \quad \neg(\forall\, i,a \in {\mathbb{A}}), \label{eq:ren_f} \\
\tens{\rh}{ij}{ab} &= 2 \tens{v}{ij}{ab} - \mpd{ij}{ab} \tens{t}{ab}{ij}. \label{eq:ren_v}
\end{align}
In Eq.~\eqref{eq:widetildeH}, we have introduced a compact notation for a string of creation and annihilation operators: $\sqop{pq\cdots}{rs\cdots} = \cop{p} \cop{q} \cdots \aop{s} \aop{r}$.
In Eq.~\eqref{eq:ren_f}, we have also defined an auxiliary one-body intermediate $\tens{\check{f}}{i}{a}$
\begin{equation}
\label{eq:fock_check}
\tens{\check{f}}{i}{a} = \tens{f}{i}{a} + \sum_{ux}^{\mathbb{A}} \mpd{x}{u} \density{x}{u} \tens{t}{ax}{iu}.
\end{equation}

The DSRG-MRPT2 cluster operator $\hat{T}$ is written as
\begin{equation}
\hat{T} = \sum_{i}^{\mathbb{H}} \sum_{a}^{\mathbb{P}} \tens{t}{a}{i} \no{ \sqop{a}{i} } + \frac{1}{4} \sum_{ij}^{\mathbb{H}} \sum_{ab}^{\mathbb{P}} \tens{t}{ab}{ij} \no{ \sqop{ab}{ij} }
\end{equation}
where the cluster amplitudes are determined via:
\begin{align}
\tens{t}{a}{i} &= \tens{\check{f}}{i}{a} \, {\cal R}_s (\mpd{i}{a}) , && \quad \neg(\forall\, i,a \in {\mathbb{A}}), \label{eq:amp_1} \\
\tens{t}{ab}{ij} &= \tens{v}{ij}{ab} \, {\cal R}_s (\mpd{ij}{ab}) , && \quad \neg(\forall\, i,j,a,b \in {\mathbb{A}}). \label{eq:amp_2}
\end{align}
Here, ${\cal R}_s(\Delta)$ is a function that regularizes the inverse of $\Delta$:
\begin{equation}
\label{eq:regulerizer}
{\cal R}_s(\Delta) = \frac{1 - e^{-s \Delta^2}}{\Delta}.
\end{equation}
As noted in Eqs.~\eqref{eq:amp_1} and \eqref{eq:amp_2}, internal excitations labeled solely by active indices are excluded from the definitions of cluster amplitudes.
For this reason, along with the fact that $\tens{f}{u}{v} (\forall\, u,v \in \mathbb{A})$ contribute only to $\hat{H}^{(0)}$, the one-body components of $\widetilde{H}$ should not include elements labeled by only active indices either [see Eq.~\eqref{eq:ren_f}].
Contrarily, no index restrictions apply to $\tens{\rh}{ij}{ab}$ [Eq.~\eqref{eq:ren_v}] because $\tens{v}{uv}{xy}$ are proper contributions to the first-order Hamiltonian.
We also point out that both $\tens{\rh}{ij}{ab}$ and $\tens{t}{ab}{ij}$ are antisymmetric with respect to individual permutations of upper or lower indices, e.g., $\tens{t}{ab}{ij} = -\tens{t}{ba}{ij} = -\tens{t}{ab}{ji} = \tens{t}{ba}{ji}$.

The denominators $\mpd{ij\cdots}{ab\cdots}$ that enter into the DSRG-MRPT2 amplitudes [Eqs.~\eqref{eq:amp_1} and \eqref{eq:amp_2}] may be positive or close to zero.
When the latter occurs, these amplitudes remain bounded for finite values of $s$ because the divergence of the denominator is suppressed by the regularizer ${\cal R}_s$ [Eq.~\eqref{eq:regulerizer}].
However, even if a denominator is not zero, the magnitude of ${\cal R}_s$ can be as large as $\approx 0.6382\sqrt{s}$.\cite{li2015multireference}
Therefore, it is necessary to use a value of $s$ that balances the amount of correlation captured by the DSRG-MRPT2 with the risk of reintroducing intruders.
Previous work\cite{li2015multireference} has found a ``Goldilocks zone'' for $s$ around $\sim 1$ \sunit that yields accurate results and avoids intruders.

\subsection{DSRG-MRPT2 Gradients}
\label{subsec:gradient}

Analytic expressions for the DSRG-MRPT2  gradients are obtained by taking the total derivatives of the energy [Eq.~\eqref{eq:energy_short}] with respect to external perturbations.
Without loss of generality, in this work we take  the external perturbations to be nuclear displacements.
The difficulty of deriving the DSRG-MRPT2 gradient theory can be easily appreciated.
Consider the derivative of an amplitude $\tens{t}{i}{a}$ with respect to an atomic coordinate $R$ in the gradient contribution $\pdev{\cal E}{\tens{t}{i}{a}} \pdev{\tens{t}{i}{a}}{R}$.
As shown in Eq.~\eqref{eq:amp_1}, all quantities that enter in the equation for $\tens{t}{i}{a}$ ($\tens{f}{a}{i}$, $\mpd{v}{u}$, and $\tens{t}{au}{iv}$) depend on $R$, leading to numerous contributions to the derivative equations.

More importantly, the DSRG-MRPT2 energy implicitly depends on both the orbital coefficients $C_{\mu p}$ [Eq.~\eqref{eq:mo}] and the CI coefficients $c_I$ [Eq.~\eqref{eq:cas_wfn}].
These quantities are determined by the CASSCF stationary conditions but are not variationally optimized in DSRG-MRPT2.
Thus, computing the DSRG-MRPT2 gradients requires the evaluation of $\pdev*{C_{\mu p}}{R}$ and $\pdev*{c_I}{R}$, which can be solved via the coupled-perturbed (CP) CASSCF equation.\cite{Osamura.1982,Yamaguchi1994book}
For a molecule with $M$ atoms, there are $3M$ CP-CASSCF equations, the solution of which becomes computationally impractical for large systems.
As realized by Handy and Schaefer,\cite{handy1984evaluation} the $3M$ CP-CASSCF equations may be replaced with a single perturbation-independent response equation (Z-vector approach), whose solution suffices to compute the energy derivatives for all nuclei.
The Lagrangian formulation of the gradient theory of Helgaker and J{\o}rgensen\cite{helgaker1988analytical}
directly leads to a set of response equations equivalent to the Z-vector approach.

Herein, we follow the standard approach for deriving analytic energy gradients based on the method of Lagrange multipliers.\cite{helgaker1988analytical,Helgaker:2002hs}
The DSRG-MRPT2 Lagrangian (${\cal L}$) reads as
\begin{align}
\label{eq:Lag_master}
{\cal L} =&\, {\cal E} + \sum_{n = 1}^{2} ( {\cal T}_n + {\cal \tilde{H}}_n) + {\cal F} + {\cal W} + {\cal X} + {\cal Y},
\end{align}
with scalar terms reflecting the constraints on the $n$-body cluster amplitudes (${\cal T}_n$), the $n$-body modified integrals (${\cal \tilde{H}}_n$), the use of semicanonical CASSCF orbitals (${\cal F}$), the orthonormality of the MSOs (${\cal W}$), and the use of a CASCI reference (${\cal X}$) subject to normalization ($\cal Y$).
In general, each of these terms is written as a dot product between a vector (or tensor) of equality constraints and the associated Lagrange \emph{multipliers},
where every constraint is a zero-valued function of some \emph{parameters}.
All terms of Eq.~\eqref{eq:Lag_master} are summarized in Table \ref{tab:notation_lagrangian} and explicit definitions are discussed in detail in section \ref{subsec:constraints}.

\begin{table}[ht!]
\caption{Summary for the DSRG-MRPT2 Lagrangian constraints.}
\label{tab:notation_lagrangian}
\ifpreprint
\renewcommand{\arraystretch}{1.25}
\begin{tabular*}{0.75\columnwidth}{@{\extracolsep{\stretch{1}}} c l @{\hspace{0.1em}} l c l @{}}
\else
\renewcommand{\arraystretch}{1.5}
\scriptsize
\begin{tabular*}{\columnwidth}{@{\extracolsep{\stretch{1}}} c l @{\hspace{0.1em}} l c >{\tiny}l @{}}
\fi
\hline
\hline
Term & \multicolumn{2}{c}{Constraints} & Multipliers & \multicolumn{1}{c}{Description} \\
\hline
${\cal T}_n$ & $\tens{T}{ab\cdots}{ij\cdots}$ & Eqs.~\eqref{eq:t1cons} \& \eqref{eq:t2cons} & $\tens{\tau}{ab\cdots}{ij\cdots}$ & $n$-body cluster amplitudes \\
${\cal \tilde{H}}_n$ & $\tens{\tilde{H}}{ij\cdots}{ab\cdots}$ & Eqs.~\eqref{eq:g1cons} \& \eqref{eq:g2cons} & $\tens{\kappa}{ij\cdots}{ab\cdots}$ & $n$-body modified integrals \\
$\cal F$ & $\tens{F}{p}{q}$ & Eqs.~\eqref{eq:fcons_cv}--\eqref{eq:fcons} & $\tens{\zeta}{p}{q}$ & CASSCF semicanonical orbitals \\
$\cal W$ & $\tens{W}{p}{q}$ & Eq.~\eqref{eq:oon_cons} & $\tens{\omega}{p}{q}$ & orthonormal orbitals \\
$\cal X$ & $X_I$ & Eq.~\eqref{eq:ci_cons} & $\xi_I$ & CI coefficients from CASCI \\
$\cal Y$ & $Y$ & Eq.~\eqref{eq:cin_cons} & $\iota$ & normalized CI coefficients \\
\hline
\hline
\end{tabular*}
\end{table}

When the Lagrangian is stationary with respect to variations of \emph{all} the parameters and multipliers, the DSRG-MRPT2 analytic energy gradients (evaluated at the reference geometry $R_0$) can be computed as:
\begin{align}
\label{eq:Ederiv}
&\left. \dev{\cal E}{R} \right|_{R=R_0} = \left. \pdev{{\cal L}}{R} \right|_{R=R_0} \notag\\
&= \sum_{pq}^{\mathbb{G}} \edens{p}{q} (\tens{h}{q}{p})^x + \sum_{pqrs}^{\mathbb{G}} \edens{pq}{rs} (\tens{v}{pq}{rs})^x + \sum_{pq}^{\mathbb{G}} \tens{\omega}{q}{p} (\tens{S}{p}{q})^x.
\end{align}
Here, $(\tens{h}{q}{p})^x$, $(\tens{v}{pq}{rs})^x$, $(\tens{S}{p}{q})^x$ are skeleton one-electron, antisymmetrized two-electron, and overlap derivative MSO integrals, respectively.\cite{Rice:1985gg,Yamaguchi1994book,levchenko2005analytic}
These quantities are multiplied by the corresponding relaxed one-body density ($\edens{p}{q}$), relaxed two-body density ($\edens{pq}{rs}$), and the energy-weighted density $\tens{\omega}{q}{p}$,
which can be obtained by collecting the respective terms in front of $\tens{h}{p}{q}$, $\tens{v}{pq}{rs}$, and $\tens{S}{p}{q}$ in $\cal L$.
Contributions to the relaxed densities are given in Appendix \ref{app:rrdm}.

\subsection{DSRG-MRPT2 Lagrangian Constraints}
\label{subsec:constraints}

\subsubsection{CASSCF Semicanonical Orbitals}

To impose that the orbitals are variationally optimized using CASSCF and satisfy the semicanonical condition [Eq.~\eqref{eq:semi}], we include the term $\cal F$ in the Lagrangian function [Eq.~\eqref{eq:Lag_master}].
This term is defined as:
\begin{equation}
{\cal F} = \sum_{pq}^{\mathbb{G}} \tens{\zeta}{p}{q} \tens{F}{p}{q},
\end{equation}
where $\tens{\zeta}{p}{q}$ are the Lagrange multipliers associated with the constraints $\tens{F}{p}{q} = 0$.
For converged CASSCF orbitals, the following conditions are satisfied:\cite{roos1980complete,werner1990comparison}
\begin{equation}
\label{eq:casscf_fock}
\tens{f}{m}{e} = 0, \quad \tens{\tilde{f}}{e}{u} = 0, \quad  \tens{f}{m}{u} - \tens{\tilde{f}}{m}{u} = 0,
\end{equation}
with the intermediate $\tens{\tilde{f}}{p}{u}$ defined by:
\begin{equation}
\label{eq:fock_tilde}
\tens{\tilde{f}}{p}{u} = \sum_{v}^{\mathbb{A}} \tens{\bar{f}}{p}{v} \density{u}{v} + \frac{1}{2} \sum_{vxy}^{\mathbb{A}} \tens{v}{pv}{xy} \density{uv}{xy}.
\end{equation}

Equation \eqref{eq:casscf_fock} is easily translated to the following constraints:
\begin{align}
\tens{F}{m}{e} &= \tens{F}{e}{m} = \tens{f}{m}{e},  && m \in \mathbb{C}, e \in \mathbb{V}  \label{eq:fcons_cv}, \\
\tens{F}{u}{e} &= \tens{F}{e}{u} = - \tens{\tilde{f}}{e}{u}, && u \in \mathbb{A}, e \in \mathbb{V} \label{eq:fcons_av}, \\
\tens{F}{u}{m} &= \tens{F}{m}{u} = \tens{f}{m}{u} - \tens{\tilde{f}}{m}{u}, && u \in \mathbb{A}, m \in \mathbb{C},\label{eq:fcons_ac}
\end{align}
where the symmetry of $\tens{F}{p}{q}$ reflects the Hermiticity of $\tens{f}{p}{q}$ and $\tens{\tilde{f}}{p}{q}$.

We impose the semicanonical orbital basis condition [Eq.~\eqref{eq:semi}] by defining the diagonal blocks of $\tens{F}{p}{q}$ as:
\begin{equation}
\label{eq:fcons}
\tens{F}{p}{q} = \tens{f}{p}{q} - \dfock{p} \kro{q}{p}, \quad \forall\, p, q \in \mathbb{O},\,\, \forall\, \mathbb{O} \in \{ \mathbb{C} , \mathbb{A} , \mathbb{V}\}.
\end{equation}
We point out that in our formulation the orbital energies ($\dfock{p}$) in Eq.~\eqref{eq:fcons} are treated as \emph{parameters} constrained to take the value of diagonal elements of the generalized Fock operator, as done in MC-QDPT2 and CASPT2 gradient theories.\cite{nakano1998analytic,Park:2019fh}
It can be easily checked that the quantities $\tens{F}{p}{q}$ implicitly depend on three sets of parameters:
1) the MSO coefficients $\bf C$ (via the one- and two-electron integrals),
2) the reference CI coefficients $\bf c$ (via the 1- and 2-pRDMs),
and 3) the MSO orbital energies $\boldsymbol \epsilon$.

\subsubsection{Cluster Amplitudes and Modified Integrals}
\label{sec:ss_amp_htilde_cons}

The DSRG-MRPT2 correlation energy $E^{(2)}$ [see Eq.~\eqref{eq:E2nd}] is a function of cluster amplitudes (${\bf t}_1, {\bf t}_2$), modified integrals ($\tilde{\bf h}_1, \tilde{\bf h}_2$), and the reference $n$-pRDMs, as shown in Appendix \ref{app:energy}.
To shift the dependence of $\bf C$ away from $E^{(2)}$, we consider both cluster amplitudes and modified integrals as parameters in the DSRG-MRPT2 Lagrangian.
This aspect is embodied in the ${\cal T}_n$ and ${\cal \tilde{H}}_n$ constraints in Eq.~\eqref{eq:Lag_master}, which are given by
\begin{align}
{\cal T}_n &= \frac{1}{(n!)^2} \sum_{ij\cdots}^{\mathbb{H}} \sum_{ab\cdots}^{\mathbb{P}} \tens{\tau}{ab\cdots}{ij\cdots} \tens{T}{ab\cdots}{ij\cdots}, \label{eq:Tn_cali} \\
{\cal \tilde{H}}_n &= \frac{1}{(n!)^2} \sum_{ij\cdots}^{\mathbb{H}} \sum_{ab\cdots}^{\mathbb{P}} \tens{\kappa}{ij\cdots}{ab\cdots} \tens{\tilde{H}}{ij\cdots}{ab\cdots}. \label{eq:Hn_cali}
\end{align}
The constraints for the one- and two-body cluster amplitudes are obtained by rearranging Eqs.~\eqref{eq:amp_1} and \eqref{eq:amp_2}:
\begin{align}
\tens{T}{a}{i} &= \tens{\check{f}}{i}{a} \, {\cal R}_s (\mpd{i}{a}) - \tens{t}{a}{i}, && \neg(\forall\, i,a \in {\mathbb{A}}), \label{eq:t1cons} \\
\tens{T}{ab}{ij} &= \tens{v}{ij}{ab} \, {\cal R}_s (\mpd{ij}{ab}) - \tens{t}{ab}{ij}, && \neg(\forall\, i,j,a,b \in {\mathbb{A}}), \label{eq:t2cons}
\end{align}
with the associated Lagrange multipliers $\tens{\tau}{a}{i}$ and $\tens{\tau}{ab}{ij}$, respectively.
Similarly, Eqs.~\eqref{eq:ren_f} and \eqref{eq:ren_v} result in constraints for the modified integrals:
\begin{align}
\tens{\tilde{H}}{i}{a} &= \tens{f}{i}{a} + \tens{\check{f}}{i}{a} - \mpd{i}{a} \tens{t}{a}{i} - \tens{\rh}{i}{a}, \quad \neg(\forall\, i,a \in {\mathbb{A}}), \label{eq:g1cons} \\
\tens{\tilde{H}}{ij}{ab} &= 2 \tens{v}{ij}{ab} - \mpd{ij}{ab} \tens{t}{ab}{ij} - \tens{\rh}{ij}{ab}, \label{eq:g2cons}
\end{align}
with the corresponding multipliers denoted as $\tens{\kappa}{i}{a}$ and $\tens{\kappa}{ij}{ab}$, respectively.
Notice again that Eqs.~\eqref{eq:t1cons}--\eqref{eq:g1cons} inherit the restrictions of indices from Eqs.~\eqref{eq:amp_1}, \eqref{eq:amp_2}, and \eqref{eq:ren_f}.
As far as the implicit dependence on parameters concerned in these constraints,
$\tens{T}{ab}{ij}$ depends on ${\bf t}_2$, $\bf C$, and ${\boldsymbol \epsilon}$, while $\tens{T}{a}{i}$ depends on ${\bf t}_1$, $\bf C$, ${\boldsymbol \epsilon}$, $\bf c$ and ${\bf t}_2$.
Compared to the  same-rank amplitude constraints, the one- and two-body constraints for modified integrals simply add additional dependences on $\tilde{\bf h}_1$ and $\tilde{\bf h}_2$, respectively.

\subsubsection{Reference CI Coefficients}

\label{sec:ci_coeff}
Next, we discuss constraints that arise from enforcing the variational condition on the reference and its normalization.
The reference wave function $\mref$ [Eq.~\eqref{eq:cas_wfn}] satisfies the eigenvalue problem:
\begin{equation}
\sum_{J}^{{\cal M}_0} \braket{\Phi_I | \hat{H} | \Phi_J} c_J = E_0 c_I, \quad \forall\, I \in {\cal M}_0,  \label{eq:cieq}
\end{equation}
subject to the normalization condition $\|\mathbf{c}\|_2^2 = \sum_{I}^{{\cal M}_0} c_I^2 = 1$.
As such, we may write out the CI constraints as a vector ($X_I$) and a scalar ($Y$) defined as
\begin{align}
X_I =&\, \braket{\Phi_I | \hat{H} | \mref} - E_0 c_I, \label{eq:ci_cons}\\
Y =&\, 1 - \sum_{I}^{{\cal M}_0} c_I^2 , \label{eq:cin_cons}
\end{align}
and associate each constraint of Eq.~\eqref{eq:ci_cons} with a multiplier $\xi_I$ and Eq.~\eqref{eq:cin_cons} with the multiplier $\iota$.
In Eq.~\eqref{eq:ci_cons}, we have used the fact that the $c_I$ coefficients are real to symmetrize the expression for $X_I$.
It is easily verified that the $X_I$ and $Y$ constraints only depend on the parameters $\bf C$ and $\bf c$.

In the DSRG-MRPT2 Lagrangian [Eq.~\eqref{eq:Lag_master}], the CI constraints are imposed via both $\cal X$ and $\cal Y$:
\begin{align}
{\cal X} &= \sum_{I}^{{\cal M}_0} \xi_I X_I
= (E^{\rm c}_0 - E_0) \sum_{I}^{{\cal M}_0} \xi_I c_I + \tilde{E}_0^{\rm a}, \label{eq:ci_Lag} \\
{\cal Y} &= \iota \Big( 1 - \sum_{I}^{{\cal M}_0} c_I^2 \Big), \label{eq:cin_Lag}
\end{align}
where $E^{\rm c}_0$ has been defined in Eq.~\eqref{eq:E0c}.
The term $\tilde{E}_0^{\rm a}$ in Eq.~\eqref{eq:ci_Lag} is similar to Eq.~\eqref{eq:E0a} except that the 1- and 2-pRDMs in Eq.~\eqref{eq:E0a} should be replaced to the corresponding modified RDMs (mRDMs) given by:
\begin{equation}
\label{eq:mrdm}
\mdensity{uv\cdots}{xy\cdots} = \sum_{IJ}^{{\cal M}_0} \xi_I c_J \braket{\Phi_I | \sqop{uv\cdots}{xy\cdots} |\Phi_J}.
\end{equation}
As noted by Celani and Werner,\cite{celani2003analytical} 
any multiple of $c_I$ can be added to $\xi_I$ without altering the Lagrangian contribution $\cal X$ (since $\sum_I^{{\cal M}_0} c_I X_I = 0$).
It is thus convenient to use this degree of freedom to make $\bf \boldsymbol{\xi}$ and $\bf c$ orthogonal:
\begin{equation}
\label{eq:cixi}
\sum_{I}^{{\cal M}_0} \xi_I c_I = 0.
\end{equation}
Given such orthogonality condition, the CI constraint $\cal{X}$ [Eq.~\eqref{eq:ci_Lag}] can be further simplified to only one term $\tilde{E}_0^{\rm a}$.

\subsubsection{Orbital Orthonormality}

Lastly, the orthonormality of MSOs is imposed via the Lagrangian term $\cal W$:
\begin{equation}
\label{eq:W_Lag}
{\cal W} = \sum_{pq}^{\mathbb{G}} \tens{\omega}{p}{q} \tens{W}{p}{q},
\end{equation}
where the constraints are defined by
\begin{equation}
\label{eq:oon_cons}
\tens{W}{p}{q} = \kro{q}{p} - \tens{S}{p}{q}.
\end{equation}
The multipliers $\tens{\omega}{p}{q}$ are identified as elements of the energy-weighted density matrix.
The MSO orthonormality constraint [Eq.~\eqref{eq:oon_cons}] depends parametrically only on the orbital coefficients $\bf C$.

\subsection{DSRG-MRPT2 Lagrange Multipliers}
\label{subsec:multipliers}

After defining each term in the DSRG-MRPT2 Lagrangian, we solve for the Lagrange multipliers by imposing stationarity with respect to all the parameters  ($\mathbf{C}, \mathbf{c}, \tilde{\bf h}_1, \tilde{\bf h}_2, \mathbf{t}_1, \mathbf{t}_2$, and ${\boldsymbol \epsilon}$).
These parameters can be separated into two categories.
The orbital and CI coefficients stationary conditions resemble the coupled perturbed CASSCF equations, which require an iterative procedure for the solution of the corresponding multipliers ($\boldsymbol \zeta$ and $\boldsymbol \xi$).
Instead, the multipliers associated with the remaining parameters can be obtained in a direct way.

\subsubsection{Modified Integrals}

We first solve the Lagrange multipliers ${\boldsymbol \kappa}_1$ and ${\boldsymbol \kappa}_2$ corresponding to the modified integrals constraints.
Taking the derivative of $\cal L$ with respect to the modified integrals and setting them to zero leads to:
\begin{align}
\frac{\partial \cal L}{\partial \tens{\rh}{i}{a}} &= 0 \quad \Rightarrow& \tens{\kappa}{i}{a} &= \frac{\partial E^{(2)}}{\partial \tens{\rh}{i}{a}} = \braket{\mref| [ \no{\sqop{i}{a}} , \hat{T}] |\mref}, \label{eq:kappa1} \\
\frac{\partial \cal L}{\partial \tens{\rh}{ij}{ab}} &= 0 \quad \Rightarrow& \tens{\kappa}{ij}{ab} &= 4 \frac{\partial E^{(2)}}{\partial \tens{\rh}{ij}{ab}} = \braket{\mref| [ \no{\sqop{ij}{ab}} , \hat{T}] |\mref}. \label{eq:kappa2}
\end{align}
We point out that 1) $\tens{\kappa}{ij}{ab}$ is antisymmetric with respect to individual permutations of upper or lower indices
and 2) those elements labeled by active indices are zero ($\tens{\kappa}{u}{v} = \tens{\kappa}{uv}{xy} = 0, \, \forall\, u,v,x,y \in {\mathbb{A}}$).
Explicit expressions of $\tens{\kappa}{i}{a}$ and $\tens{\kappa}{ij}{ab}$ are reported in Appendix \ref{app:kappa_tau}, where we evaluate the fully connected terms of the commutators in Eqs.~\eqref{eq:kappa1} and \eqref{eq:kappa2}.
Identical expressions can be alternatively obtained by directly taking the partial derivatives of $E^{(2)}$ with respect to the modified integrals ($\tens{\rh}{i}{a}$ and $\tens{\rh}{ij}{ab}$) and antisymmetrizing the resulting contributions to $\tens{\kappa}{ij}{ab}$ with respect to index permutations.

\subsubsection{Cluster Amplitudes}

The Lagrange multipliers for the one-body cluster amplitudes can be easily solved:
\begin{equation}
\label{eq:tau1}
\frac{\partial \cal L}{\partial \tens{t}{a}{i}} = 0
\quad \Rightarrow \quad
\tens{\tau}{a}{i} = \frac{\partial E^{(2)}}{\partial \tens{t}{a}{i}} - \tens{\kappa}{i}{a} \mpd{i}{a}.
\end{equation}
For the two-body multipliers, we have
\begin{equation}
\label{eq:tau2}
\frac{\partial \cal L}{\partial \tens{t}{ab}{ij}} = 0 \, \Rightarrow \, \tens{\tau}{ab}{ij} = 4 \frac{\partial}{\partial \tens{t}{ab}{ij}} \big( E^{(2)} + {\cal T}_1 + {\cal \tilde{H}}_1 \big) - \tens{\kappa}{ij}{ab} \mpd{ij}{ab}.
\end{equation}
The derivatives of the second-order energy correction with respect to cluster amplitudes can be written as:
\begin{align}
\frac{\partial E^{(2)}}{\partial \tens{t}{a}{i}} &= \braket{\mref| [ \widetilde{H}, \no{\sqop{a}{i}}] |\mref}, \label{eq:E2nd_deriv_t1} \\
\frac{\partial E^{(2)}}{\partial \tens{t}{ab}{ij}} &= \frac{1}{4} \braket{\mref| [ \widetilde{H}, \no{\sqop{ab}{ij}}] |\mref}, \label{eq:E2nd_deriv_t2}
\end{align}
and their explicit expressions are presented in Appendix \ref{app:kappa_tau}.
To continue, we evaluate the partial derivatives of $\tens{\check{f}}{k}{c}$ [Eq.~\eqref{eq:fock_check}] with respect to $\tens{t}{ab}{ij}$:
\begin{equation}
\label{eq:deriv_fcheck_t2}
\pdev{\tens{\check{f}}{k}{c}}{\tens{t}{ab}{ij}} = \frac{1}{4} {\cal P} (ab) {\cal P} (ij) ( \mpd{a}{i} \density{a}{i} \kro{c}{b} \kro{j}{k} ),
\end{equation}
where ${\cal P}(pq)$ is an antisymmetrizer with respect to indices $p$ and $q$: ${\cal P} (pq) f(p,q,r,\dots) = f(p,q,r,\dots) - f(q,p,r,\dots)$.
We may then calculate the ${\cal T}_1$ and ${\cal \tilde{H}}_1$ terms in Eq.~\eqref{eq:tau2} as
\begin{align}
\frac{\partial {\cal T}_1}{\partial \tens{t}{ab}{ij}}
&= \frac{1}{4} {\cal P} (ab) {\cal P} (ij) \big[ \mpd{a}{i} \density{a}{i} \tens{\tau}{b}{j} \, {\cal R}_s (\mpd{j}{b}) \big], \label{eq:tau1_dt2} \\
\frac{\partial {\cal \tilde{H}}_1}{\partial \tens{t}{ab}{ij}} 
&= \frac{1}{4} {\cal P} (ab) {\cal P} (ij) \big( \mpd{a}{i} \density{a}{i} \tens{\kappa}{j}{b} \big). \label{eq:kappa1_dt2}
\end{align}
Two aspects are worth mentioning.
First, only the active-active block of the 1-pRDM contributes to Eqs.~\eqref{eq:deriv_fcheck_t2}--\eqref{eq:kappa1_dt2}.
Hence, $\density{a}{i}$ may be replaced with $\density{u}{v}$ ($\forall\, u,v \in \mathbb{A}$) after appropriate reindexing.
Second, multipliers labeled only by active indices ($\tens{\tau}{u}{v}$ and $\tens{\tau}{uv}{xy}$, $\forall\, u,v,x,y \in {\mathbb{A}}$) are not defined because internal excitations are forbidden and they are conveniently set to zero in our implementation.

\subsubsection{Orbital Energies}

The diagonal elements of $\boldsymbol \zeta$ can be obtained by making the Lagrangian stationary with respect to the semicanonical orbital energies:
\begin{equation}
\label{eq:zeta_diag}
\pdev{\cal L}{\dfock{p}} = 0 \quad \Rightarrow \quad \tens{\zeta}{p}{p} = \pdev{}{\dfock{p}} \Big[ \sum_{n = 1}^{2} ( {\cal T}_n + {\cal \tilde{H}}_n ) \Big] .
\end{equation}
Evaluating the derivatives that enter into Eq.~\eqref{eq:zeta_diag} is straightforward and the resulting expressions are provided in Appendix \ref{app:zeta_diag}.

\subsubsection{Energy-Weighted Density, Orbital Rotations, and CI Coefficients}

The remaining unknowns are the energy-weighted density ($\boldsymbol \omega$) and the orbital ($\boldsymbol \zeta$) and CI ($\boldsymbol \xi$ and $\iota$) multipliers.
In principle, these quantities are all coupled together, but as shown by Celani and Werner,\cite{celani2003analytical} it is possible to write separate equations for $\boldsymbol \zeta$ and $\boldsymbol \xi$ from those for $\boldsymbol \omega$. 
The equations for $\boldsymbol \zeta$ and $\boldsymbol \xi$ form a coupled linear systems, whose solution may be then used to evaluate $\boldsymbol \omega$.

When differentiating the Lagrangian with respect to the orbital coefficients $\bf C$, it is convenient to express this quantity as a unitary transformation of the unperturbed orbitals (${\bf C}_0$):

\begin{equation}
{\bf C} = {\bf C}_0 \exp(\boldsymbol \vartheta).
\end{equation}
Here, $\boldsymbol \vartheta$ is an anti-Hermitian matrix whose elements become the actual variational parameters.
This parameterization ensures that the perturbed orbitals remain orthonormal. 

Imposing the stationarity of the Lagrangian with respect to orbital rotations
\begin{equation}
\label{eq:zeta}
\left( \pdev{\cal L}{\boldsymbol \vartheta} \right)_{\boldsymbol \vartheta = 0}
= \left( {\bf C}^\dag \pdev{\cal L}{\bf C} \right)_{\boldsymbol \vartheta = 0} = 0,
\end{equation}
yields a set of equations that depend on $\boldsymbol \omega$, $\boldsymbol \zeta$, and $\boldsymbol \xi$ (via the mRDMs $\mdensity{uv\cdots}{xy\cdots}$), as reported in Appendix \ref{app:zeta}.
A way to decouple $\boldsymbol \omega$ from the other variables is suggested by the structure of the MSO overlap contribution to Eq.~\eqref{eq:zeta} 
\begin{equation}
\label{eq:overlab_orb}
\pdev{\cal W}{\tens{\vartheta}{p}{q}} = \pdev{\cal W}{\tens{\vartheta}{q}{p}}
= - \sum_{r} ( \tens{\omega}{p}{r} \tens{S}{q}{r} + \tens{\omega}{r}{p} \tens{S}{r}{q} )
= -( \tens{\omega}{p}{q} + \tens{\omega}{q}{p} ).
\end{equation}
To remove the dependence on $\boldsymbol \omega$, it is sufficient to consider the antisymmetric part of $\pdev*{\cal L}{\tens{\vartheta}{p}{q}}$,
\begin{equation}
\label{eq:asym_lag_orbs}
\pdev{\cal L}{\tens{\vartheta}{p}{q}} - \pdev{\cal L}{\tens{\vartheta}{q}{p}} = 0,
\end{equation}
which only depends on the unsolved orbital ($\boldsymbol \zeta$) and CI ($\boldsymbol \xi$) multipliers.
Equation~\eqref{eq:asym_lag_orbs} forms a set of linear equations of the form 
\begin{equation}
\label{eq:orb_response}
{\bf A}^{\rm oo} {\boldsymbol \zeta} + {\bf A}^{\rm oc} {\boldsymbol \xi} = {\bf b}^{\rm o},
\end{equation}
where ${\bf A}^{\rm oo}$ and ${\bf A}^{\rm oc}$ are matrices of dimension $N_\mathrm{indep}^2$  and $N_\mathrm{indep} N_\mathrm{det}$, where 
$N_\mathrm{indep}$ is the number of independent orbital rotation parameters and $N_\mathrm{det}$ the number of CI determinants.
The vector $ {\bf b}^{\rm o}$ collects all constant terms and is of dimension $N_\mathrm{indep}$.
Equation~\eqref{eq:orb_response} alone is insufficient to determine $\boldsymbol \zeta$ and $\boldsymbol \xi$, and must be augmented with additional conditions obtained from imposing stationarity with respect to the CI coefficients.

The derivative of $\mathcal{L}$ with respect to the CI coefficients takes the form
\begin{equation}
\label{eq:xi}
\pdev{\cal L}{c_I} = \pdev{}{c_I} ({\cal E + T_{\rm 1} + \tilde{H}_{\rm 1} + F + X + Y}) = 0, \quad\forall I \in {\cal M}_0.
\end{equation}
Equation \eqref{eq:xi} consists of a large set of linear equations for the orbital multipliers $\boldsymbol \zeta$ and the CI multipliers $\boldsymbol \xi$, that is, 
\begin{equation}
\label{eq:ci_response}
{\bf A}^{\rm co} {\boldsymbol \zeta} + {\bf A}^{\rm cc} {\boldsymbol \xi} = {\bf b}^{\rm c},
\end{equation}
where the matrices ${\bf A}^{\rm co}$ and ${\bf A}^{\rm cc}$ are of size $N_\mathrm{indep} N_\mathrm{det}$ and $N_\mathrm{det}^2$, respectively, while the vector ${\bf b}^{\rm c}$ contains $N_\mathrm{det}$ entries.
The Lagrange multiplier connected to the CI normalization condition [$\iota$, see Eq.~\eqref{eq:cin_Lag}] can be computed as:
\begin{equation}
\label{eq:iota}
\iota = \frac{1}{2} \sum_{I}^{{\cal M}_0} c_I \pdev{}{c_I} ( {\cal E + T_{\rm 1} + \tilde{H}_{\rm 1} + F} ),
\end{equation}
which depends on the orbital multipliers $\boldsymbol \zeta$ (see Appendix \ref{app:xi}).
Equation \eqref{eq:iota} is obtained from Eq.~\eqref{eq:xi} ($\sum_{I}^{{\cal M}_0} c_I \pdev{\cal L}{c_I} = 0$) using the fact that $\|\mathbf{c}\|_2^2 = 1$ and $\hat{H} \ket{\mref} = E_0 \ket{\mref}$.

The linear equations for the orbital and CI multipliers [Eqs.~\eqref{eq:orb_response} and \eqref{eq:ci_response}] may be combined into a single linear system of the form ${\bf A x} = {\bf b}$ with entries defined as follows
\begin{align}
\bf A \equiv
\begin{pmatrix}
{\bf A}^{{\rm oo}} & {\bf A}^{{\rm oc}} \\
{\bf A}^{{\rm co}} & {\bf A}^{{\rm cc}}
\end{pmatrix},
\quad
\bf x \equiv
\begin{pmatrix}
\boldsymbol \zeta \\
\boldsymbol \xi
\end{pmatrix},
\quad
\bf b \equiv
\begin{pmatrix}
{\bf b}^{{\rm o}} \\
{\bf b}^{{\rm c}}
\end{pmatrix}.
\label{eq:linearaxb}
\end{align} 
When written in this form, ${\bf A}$ may be identified as a Jacobian matrix.
Expressions for all the blocks of ${\bf A}$ and ${\bf b}$ are reported in Appendix \ref{app:cpcasscf}.
We postpone the discussion of how this linear system is solved to Sec.~\ref{sec:iterative_solution}.

Once $\boldsymbol \zeta$ and $\boldsymbol \xi$ are determined, the symmetric counterpart of Eq.~\eqref{eq:asym_lag_orbs} (i.e., $\pdev*{\cal L}{\tens{\vartheta}{p}{q}} + \pdev*{\cal L}{\tens{\vartheta}{q}{p}} = 0$) can be used to recover the energy-weighted density:
\begin{align}
\tens{\omega}{p}{q} = \frac{1}{4} \big( \pdev{}{\tens{\vartheta}{p}{q}} + \pdev{}{\tens{\vartheta}{q}{p}} \big) \Big[ E_0 + & \sum_{n = 1}^{2} ( {\cal T}_n + {\cal \tilde{H}}_n) + {\cal F} + {\cal X} \Big].
\end{align}
Here, we have ignored the contribution from $E^{(2)}$ because it is independent on orbital rotations, as stated in Sec.~\ref{sec:ss_amp_htilde_cons}.

\subsubsection{Iterative Solution of the Orbital and CI Multipliers}

\label{sec:iterative_solution}

The system of linear equations for the orbital and CI multipliers involve $N_\mathrm{indep} + N_\mathrm{det}$ variables.
When either the number of independent pairs or the size of the CI space becomes too large, it is unfeasible to store ${\bf A}$ explicitly and solve the linear system by direct inversion.
The standard solution to this problem is employ a direct iterative linear solver that directly builds a vector ${\boldsymbol \sigma} \equiv {\bf Ax}$, thus avoiding the storage problem.

As noted in Sec.~\ref{sec:ci_coeff}, to find a unique solution to the CI multiplier equations, we impose the constraint 
${\boldsymbol \xi} \cdot {\bf c} = 0$ [see Eq.~\eqref{eq:cixi}].
To enforce this constraint in the solution of the linear system, we define a projection matrix $\bf P$
\begin{equation}
{\bf P} = 
\begin{pmatrix}
{\bf 1} & {\bf 0} \\
{\bf 0} & {\bf 1} - {\bf c c}^T
\end{pmatrix}.
\end{equation}
Then the constraint ${\boldsymbol \xi} \cdot {\bf c} = 0$ is equivalent to the condition $\bf Px = x$, and we may use this result to write the linear system in the form
\begin{equation}
\bf (PAP) x = Pb.
\end{equation}
The matrix ${\bf PAP}$ is rank deficient since the vector ${\bf x}^\parallel = {(\bf 0, c)}^T$ is such that $\bf Px^\parallel = 0$.
This linear system can be solved using the generalized minimal residual method without explicitly storing the matrix $\bf PAP$.

\subsection{Computational Cost}
\label{subsec:impl}

We end this section by briefly discussing the computational cost of the DSRG-MRPT2 analytic gradients.
In general, the cost of solving the Lagrange multipliers has the same scaling of a DSRG-MRPT2 single-point computation with a slightly larger prefactor.
A vanilla DSRG-MRPT2 energy computation based on the CASSCF orbitals can be largely separated into four steps:
\begin{enumerate}
\item Solve the CASSCF problem for the orbital and CI coefficients.
\item Compute the 1-, 2-, and 3-pRDMs using the CASSCF wave function.
\item Transform the one- and two-electron integrals to the MO basis.
\item Build the modified integrals and cluster amplitudes, and use these quantities to evaluate the DSRG-MRPT2 correlation energy via tensor contractions.
\end{enumerate}
In comparison, the gradient computations take the following additional steps:
\begin{enumerate}
\setcounter{enumi}{4}
\item Compute the multipliers ${\boldsymbol \kappa}_1, {\boldsymbol \kappa}_2, {\boldsymbol \tau}_1, {\boldsymbol \tau}_2$, and $\tens{\zeta}{p}{p}$.
\item Setup the coupled linear system [Eq.~\eqref{eq:linearaxb}] and solve for the multipliers $\boldsymbol \zeta$ and $\boldsymbol \xi$.
\item Form the relaxed density matrices and energy-weighted density.
\item Transform the MO densities in step 7 to the AO basis and contract it with skeleton derivative integrals.
\end{enumerate}

We can see a rough correspondence between these two procedures.
For example, step 5 in the gradients computation corresponds to step 4 of the energy computation.
In these two steps, with the assumption of using a small active space ($N_\mathrm{A} \ll N_\mathrm{C} < N_\mathrm{V}$), the computational cost is dominated by the tensor contraction of an MP2-like term with a scaling of ${\cal{O}}(N_{\rm C}^2 N_{\rm V}^2)$.
The cost of solving the linear system (step 6) is slightly higher than the cost of second-order CASSCF optimization (step 1).
In fact, the linear system [Eq.~\eqref{eq:linearaxb}] is analogous to the Newton optimization step in CASSCF, where $\bf A$ and $\bf b$ correspond to the Hessian and the gradient vector, respectively.
However, the ${\bf b}^{\rm c}$ vector contains terms involving $\pdev*{\density{uvw}{xyz}}{c_I}$, which share the same ${\cal{O}}(N_{\rm A}^6 N_{\rm det})$ scaling of computing the 3-pRDM.
This steep cost may be avoided by considering the additional tensor contractions and introducing clever intermediates, as suggested in Ref.~\citenum{chatterjee2020extended}.
The computational scaling for the MO to AO transformation of the relaxed densities (step 8) is identical to the integral transformation step for the DSRG-MRPT2 energy (step 3).
Overall, we see that the computational cost to obtain the analytical gradients is similar to that of an energy computation.

\section{Results}
\label{sec:results}

We implemented the DSRG-MRPT2 analytic energy gradients in the open-source program \forte.\cite{FORTE2021}
The one- and two-electron integrals along with the corresponding derivative integrals were obtained from \PSI 1.4.\cite{smith2020psi4}
The correctness of the implementation was validated by comparing the analytic gradients against five-point finite-difference numerical gradients using a 0.005 a.u. step size.
In particular, we tested the gradient and the optimized bond lengths of \ce{HF} and \ce{N2} using CASSCF(2,2) and CASSCF(6,6) reference wave functions, respectively.
The cc-pCVDZ basis set\cite{dunning1989a,woon1995a} was used for all computations in this work.

As a pilot application of the DSRG-MRPT2 gradient theory, we optimized the geometry of {\it p}-benzyne for both singlet and triplet states.
The resulting geometries were used to compute the adiabatic singlet--triplet gap ($\Delta E_{\rm ST} = E_{\rm T} - E_{\rm S}$).
We compared the DSRG-MRPT2 results to those of CASPT2,\cite{werner1996third} the partially contracted NEVPT2 (pc-NEVPT2),\cite{angeli2001introduction} and Mukherjee's state-specific multireference coupled cluster theory with singles, doubles, and perturbative triples [Mk-MRCCSD(T)].\cite{mahapatra1998state,evangelista2010perturbative}
We employed the minimal CAS(2,2) active space that consists of two electrons in the two $\sigma$ orbitals located on the dehydrogenated carbon atoms.
For geometry optimizations, the maximum component of the gradient was converged to less than $2 \times 10^{-6}$ a.u.
Both CASPT2 and pc-NEVPT2 results were obtained using \molpro 2015.1\cite{MOLPRO2015} while those from Mk-MRCCSD(T) were computed using \PSI 1.4.\cite{smith2020psi4}

\begin{figure}[ht!]
\centering
\ifpreprint
\includegraphics[width=0.5\columnwidth]{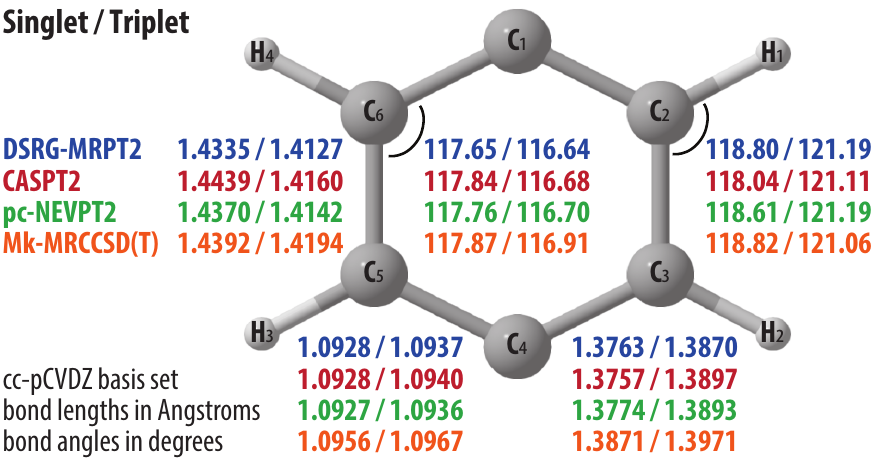}
\else
\includegraphics[width=1.0\columnwidth]{figure2}
\fi
\caption{Equilibrium geometries of singlet and triplet \textit{p}-benzyne optimized using various multireference methods using the cc-pCVDZ basis set. The DSRG flow parameter was set to 1.0\sunit. The CASSCF(2,2) reference was used for all computations.}
\label{fig:benzyne_comparison}
\end{figure}

Figure \ref{fig:benzyne_comparison}
 presents the DSRG-MRPT2 optimized geometries of singlet and triplet \textit{p}-benzyne.
Here, we set the flow parameter to $s = 1.0$ \sunit, a value that previously shown to yield reliable singlet--triplet gap of \textit{p}-benzyne.\cite{li2015multireference}
The DSRG-MRPT2 optimized geometries are in excellent agreements to those of CASPT2 and pc-NEVPT2.
For example, the DSRG-MRPT2 bond lengths and angles deviate from those of CASPT2 by at most 1.0 pm (\ce{C5-C6} of the singlet) and $0.8^\circ$ (\ce{\angle H1C2C3} of the singlet), respectively.
Compared to Mk-MRCCSD(T), DSRG-MRPT2  underestimates all \ce{C-C} bonds by roughly 1 pm for both singlet and triplet states.

\begin{figure}[ht!]
\centering
\ifpreprint
\includegraphics[width=0.5\columnwidth]{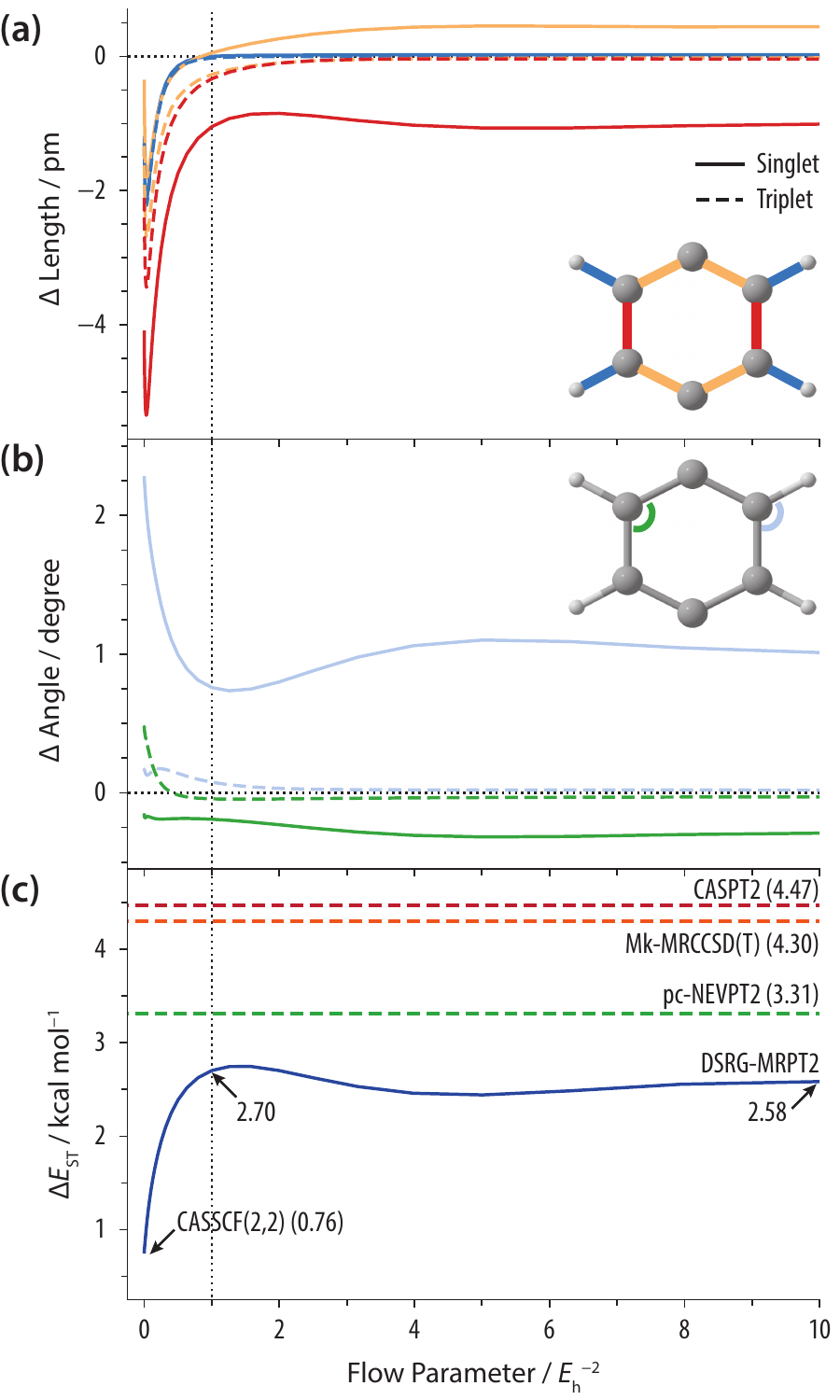}
\else
\includegraphics[width=1.0\columnwidth]{figure3}
\fi
\caption{(a) Bond lengths and (b) bond angles from DSRG-MRPT2 optimized geometries relative to the CASPT2 optimized geometries as a function of the flow parameter. (c) DSRG-MRPT2 adiabatic singlet--triplet gap as a function of the flow parameter.}
\label{fig:s_dependence}
\end{figure}

In Fig.~\ref{fig:s_dependence}, we show the sensitivity of the DSRG-MRPT2 optimized bond distances and angles [computed as deviations from CASPT2 values] with respect to the flow parameter $s$.
As $s$ increases from 0, all geometric parameters vary quickly and converge roughly around $s = 1$ \sunit.
Interestingly, the bond lengths first decrease for $s < 0.04$ \sunit and then start to increase for $0.04 < s < 1$ \sunit.
When $s$ keeps growing from 1 to 10 \sunit, small yet noticeable changes are observed for the \ce{C5-C6} bond ($\leq 0.4$ pm) and \ce{\angle H1C2C3} ($< 0.4^\circ$) of the singlet.

The bottom panel of Fig.~\ref{fig:s_dependence} also reports the adiabatic singlet--triplet splittings of {\it p}-benzyne computed using various multireference methods.
For DSRG-MRPT2, we again see a quick convergence of $\Delta E_{\rm ST}$ near $s = 1$ \sunit, while further increase of $s$ leads to only 0.1 \kcal difference in $\Delta E_{\rm ST}$.
The DSRG-MRPT2 ($s = 1$) prediction of $\Delta E_{\rm ST}$ is 2.7 \kcal, which is 0.6, 1.8, 1.6 \kcal smaller than that of pc-NEVPT2, CASPT2, and Mk-MRCCSD(T), respectively.
This underestimation has been observed previously, even when $\Delta E_{\rm ST}$ is computed using optimized Mk-MRCCSD/cc-pVTZ geometries.\cite{li2015multireference}
However, we note that the $\Delta E_{\rm ST}$ of DSRG-MRPT2 can be improved via reference relaxation.\cite{li2017driven}
Specifically, using the corresponding unrelaxed DSRG-MRPT2 geometries reported in Fig.~\ref{fig:benzyne_comparison}, the partially relaxed and the relaxed versions of DSRG-MRPT2 ($s = 1$) predict the $\Delta E_{\rm ST}$ to be 3.57 and 3.76 \kcal, respectively.
These values fall in between the pc-NEVPT2 and Mk-MRCCSD(T) results.

\section{Conclusions}
\label{sec:conclusions}

We have derived and implemented the analytic energy gradients of the unrelaxed DSRG-MRPT2 approach.
Our derivation uses the method of Lagrange multipliers to impose constraints on the first-order modified integrals and cluster amplitudes, the orbitals, and CI coefficients.
Despite the complexity of the final Lagrangian, analytic expressions for the DSRG-MRPT2 gradient could still be derived by hand (albeit via a laborious procedure).
Inheriting the numerical robustness of the DSRG equations, the corresponding Lagrangian is similarly well-behaved even when small energy denominators arise, circumventing the intruder state-problem in linear-response computations.

We have used the DSRG-MRPT2 analytic gradients to optimize the equilibrium structures of {\it p}-benzyne and study the dependence of the optimized geometry on the flow parameter $s$.
The optimized geometries of both the singlet and triplet states show very good agreement with those computed with CASPT2, pc-NEVPT2, and Mk-MRCCSD(T).
Comparing geometries between the two states, those of the singlet are more sensitive to the value of $s$. The $s$-dependency plot also shows that sufficient correlation contributions are captured with $s$ equal to $1.0$ \sunit.
Finally, we investigate the singlet-triplet splittings of {\it p}-benzyne, which are underestimated using the DSRG-MRPT2 formalism compared against other MRPT2 approaches. 

The current work paves a way for expanding the application of MR-DSRG methods beyond the computation of energies.
When compared to other MRPT2 methods, one significant advantage of the DSRG-MRPT2 is the intrinsically lower computational cost due to the absence of the 4-pRDM in the formalism.
The cost of DSRG-MRPT2 may be further reduced by neglecting the 3-body cumulants, without introducing artificial intruders.\cite{Zgid:2009fu,li2015multireference}
With this approximation, one avoids the computation of the reference 3-pRDM.
The active space dependence of the DSRG-MRPT2 gradients is reduced to ${\cal{O}}(N_{\rm A}^5 N_{\rm V})$ for the correlation energy terms and ${\cal{O}}(N_{\rm A}^4 N_{\rm det})$ for the CASCI contribution to the response equations.
To test the accuracy of this approximation, we re-optimized the singlet and triplet geometries of \textit{p}-benzyne using $s=1$ \sunit.
Comparing to the geometries optimized using the complete DSRG-MRPT2 theory (i.e., with 3-body cumulant contributions), the equilibrium bond lengths and bond angles deviate by at maximum 0.4 pm (\ce{C5-C6} of the singlet) and 0.5$^\circ$ (\ce{\angle H1C2C3} of the singlet), respectively.
We also employed these geometries yet computed the $\Delta E_{\rm ST}$ using the complete DSRG-MRPT2 theory.
The resulting spin gap is 2.67 \kcal, only 0.03 \kcal smaller than the exact answer (see Fig.~\ref{fig:s_dependence}).
Such negligible deviation indicates that the pruned DSRG-MRPT2 scheme may be used to optimize the geometry without significant degradation of the accuracy.

The current results motivate us for further developments of the theory.
On the algorithmic side, an important bottleneck is the high memory cost required to store the two-electron integrals, which may be alleviated by applying resolution the identity techniques (i.e., density fitting).
Another important limitation of the current implementation is the explicit storage of derivatives of the reduced density matrices required to solve the coupled equations for the orbital and CI multipliers.
From the theory perspective, it would highly desirable to develop the gradient theory for the state-averaged DSRG-MRPT2\cite{Li:2018kl} to allow optimizations of both ground- and excited-state PESs, in particular near conical intersections.
A more challenging future extension is the development of analytic gradients of higher-order MR-DSRG theories, including third-order perturbation theory and nonperturbative MR-DSRG methods.
These extensions would require significant human effort and the use of automatic implementation techniques.\cite{abbott2021arbitrary,macleod2015communication,Song2021}

\appendix

\titleformat{\section}
{\normalfont\large\sffamily\bfseries\color{goodred}}
{\raisebox{5pt}{\colorbox{goodred}{\rule[-2pt]{2pt}{2pt}}}}{0.5em}{APPENDIX \thesection.~\uppercase}

\setcounter{equation}{0}
\renewcommand{\theequation}{A\arabic{equation}}

\setcounter{table}{0}
\renewcommand{\thetable}{A\arabic{table}}

\section{DSRG-MRPT2 Energy Expression}
\label{app:energy}

The DSRG-MRPT2 energy contributions are reported in Table \ref{tab:energy_rdm}.
In our previous work,\cite{li2015multireference,hannon2016integral} these terms are written in terms of cluster amplitudes, modified first-order integrals, 1-pRDM, one-hole RDM (1-hRDM), and two- and three-body density cumulants.
In contrast, here we expand the 1-hRDM and all density cumulants in terms of 1-, 2-, and 3-pRDMs for the purpose of deriving the CI response terms (see Appendix \ref{app:xi}).

\begin{table*}[h!]
\ifpreprint
\renewcommand{\arraystretch}{1.25}
\else
\renewcommand{\arraystretch}{1.35}
\fi
\caption{DSRG-MRPT2 energy contributions expressed in terms of the modified first-order integrals ($\rh$), the first-order cluster amplitudes ($t$), and the reference 1-, 2-, and 3-pRDMs ($\gamma$). More compact expressions can be found in Ref.~\citenum{li2015multireference}.}
\label{tab:energy_rdm}
\begin{tabular*}{\textwidth}{@{\extracolsep{\stretch{1}}} l l@{}}
\hline
\hline

Term & Energy Expression \\

\hline

A1 & $+ \sum_{a}^{\mathbb{P}} \sum_{m}^{\mathbb{C}} \tens{\rh}{m}{a} \tens{t}{a}{m}$ \\
A2 & $+ \sum_{e}^{\mathbb{V}} \sum_{uv}^{\mathbb{A}} \tens{\rh}{v}{e} \tens{t}{e}{u} \density{v}{u} - \sum_{m}^{\mathbb{C}} \sum_{uv}^{\mathbb{A}} \tens{\rh}{m}{u} \tens{t}{v}{m} \density{v}{u}$ \\

\hline

B1 & $+ \frac{1}{2} \sum_{e}^{\mathbb{V}} \sum_{uvxy}^{\mathbb{A}} \tens{\rh}{x}{e} \tens{t}{ey}{uv} \density{xy}{uv} - \frac{1}{2} \sum_{m}^{\mathbb{C}} \sum_{uvxy}^{\mathbb{A}} \tens{\rh}{m}{v} \tens{t}{xy}{um} \density{xy}{uv}$ \\
B2 & $- \sum_{e}^{\mathbb{V}} \sum_{uvxy}^{\mathbb{A}} \tens{\rh}{x}{e} \tens{t}{ey}{uv} \density{x}{u} \density{y}{v} + \sum_{m}^{\mathbb{C}} \sum_{uvxy}^{\mathbb{A}} \tens{\rh}{m}{v} \tens{t}{xy}{um} \density{x}{u} \density{y}{v}$ \\

\hline

C1 & $+ \frac{1}{2} \sum_{e}^{\mathbb{V}} \sum_{uvxy}^{\mathbb{A}} \tens{\rh}{xy}{ev} \tens{t}{e}{u} \density{xy}{uv} - \frac{1}{2} \sum_{m}^{\mathbb{C}} \sum_{uvxy}^{\mathbb{A}} \tens{\tilde{h}}{my}{uv} \tens{t}{x}{m} \density{xy}{uv}$
 \\
C2 & $- \sum_{e}^{\mathbb{V}} \sum_{uvxy}^{\mathbb{A}} \tens{\rh}{xy}{ev} \tens{t}{e}{u} \density{x}{u} \density{y}{v} + \sum_{m}^{\mathbb{C}} \sum_{uvxy}^{\mathbb{A}} \tens{\tilde{h}}{my}{uv} \tens{t}{x}{m} \density{x}{u} \density{y}{v}$ 
\\

\hline

D1 & $+ \frac{1}{4} \sum_{ab}^{\mathbb{P}} \sum_{mn}^{\mathbb{C}} \tens{\rh}{mn}{ab} \tens{t}{ab}{mn}$ \\
D2 & $+ \frac{1}{2} \sum_{ab}^{\mathbb{P}} \sum_{m}^{\mathbb{C}} \sum_{uv}^{\mathbb{A}} \tens{\rh}{mu}{ab} \tens{t}{ab}{mv} \density{u}{v} - \frac{1}{2} \sum_{a}^{\mathbb{P}} \sum_{mn}^{\mathbb{C}} \sum_{uv}^{\mathbb{A}} \tens{\rh}{mn}{av} \tens{t}{au}{mn} \density{u}{v}$ \\

D3 & $+ \frac{1}{8} \sum_{ab}^{\mathbb{P}} \sum_{uvxy}^{\mathbb{A}} \tens{\rh}{xy}{ab} \tens{t}{ab}{uv} \density{xy}{uv} + \frac{1}{8} \sum_{mn}^{\mathbb{C}} \sum_{uvxy}^{\mathbb{A}} \tens{\rh}{mn}{uv} \tens{t}{xy}{mn} \density{xy}{uv} + \sum_{a}^{\mathbb{P}} \sum_{m}^{\mathbb{C}} \sum_{uvxy}^{\mathbb{A}} \tens{\rh}{mx}{au} \tens{t}{ay}{mv} \density{xy}{uv}$ \\
D4 & $- \sum_{a}^{\mathbb{P}} \sum_{m}^{\mathbb{C}} \sum_{uvxy}^{\mathbb{A}} \tens{\rh}{mx}{au} \tens{t}{ay}{mv} \density{x}{u} \density{y}{v}$ \\

D5 & $- \frac{1}{4} \sum_{e}^{\mathbb{V}} \sum_{uvwxyz}^{\mathbb{A}} \tens{\rh}{xy}{ew} \tens{t}{ez}{uv} \density{xyz}{uvw} + \frac{1}{4} \sum_{m}^{\mathbb{C}} \sum_{uvwxyz}^{\mathbb{A}} \tens{\rh}{mz}{uv} \tens{t}{xy}{mw} \density{xyz}{uvw}$ \\
D6 &
$
+ \frac{1}{2} \sum_{e}^{\mathbb{V}} \sum_{uvwxyz}^{\mathbb{A}} \tens{\rh}{xy}{ew} \tens{t}{ez}{uv} \density{x}{w} \density{yz}{uv}
+ \frac{1}{2} \sum_{e}^{\mathbb{V}} \sum_{uvwxyz}^{\mathbb{A}} \tens{\rh}{xy}{ew} \tens{t}{ez}{uv} \density{z}{u} \density{xy}{vw}
$ \\
D6 &
$
- \frac{1}{2} \sum_{m}^{\mathbb{C}} \sum_{uvwxyz}^{\mathbb{A}} \tens{\rh}{mz}{uv} \tens{t}{xy}{mw} \density{x}{w} \density{yz}{uv}
- \frac{1}{2} \sum_{m}^{\mathbb{C}} \sum_{uvwxyz}^{\mathbb{A}} \tens{\rh}{mz}{uv} \tens{t}{xy}{mw} \density{z}{u} \density{xy}{vw}
$ \\
D7 & $- \sum_{e}^{\mathbb{V}} \sum_{uvwxyz}^{\mathbb{A}} \tens{\rh}{xy}{ew} \tens{t}{ez}{uv} \density{y}{u} \density{z}{v} \density{x}{w} + \sum_{m}^{\mathbb{C}} \sum_{uvwxyz}^{\mathbb{A}} \tens{\rh}{mz}{uv} \tens{t}{xy}{mw} \density{y}{u} \density{z}{v} \density{x}{w}$ \\

\hline
\hline
\end{tabular*}
\end{table*}

\section{Contributions to the Relaxed Density Matrices}
\label{app:rrdm}

We first focus on the one-body relaxed density matrix elements $\tens{\Gamma}{p}{q}$.
Assuming the use of real orbitals, the nonzero elements of $\tens{\Gamma}{p}{q}$ are given by
\begin{align}
\edens{e}{f} &= \tens{\zeta}{f}{e}, && \forall\, e,f \in \mathbb{V} \label{eq:1rrdm_vv} \\
\edens{m}{n} &= \kro{m}{n} + \tens{\zeta}{n}{m}, && \forall\, m,n \in \mathbb{C} \label{eq:1rrdm_cc} \\
\edens{u}{v} &= \density{v}{u} + \mdensity{v}{u} + \tens{\zeta}{v}{u}, && \forall\, u,v \in \mathbb{A} \label{eq:1rrdm_aa} \\
\edens{e}{m} &= \tens{\alpha}{m}{e} + 2 \tens{\zeta}{m}{e}, && \forall\, e \in \mathbb{V}, \forall\, m \in \mathbb{C} \label{eq:1rrdm_vc} \\
\edens{u}{m} &= \tens{\alpha}{m}{u} + 2 \sum_{v}^{\mathbb{A}} \tens{\zeta}{m}{v} \cdensity{v}{u}, && \forall\, u \in \mathbb{A}, \forall\, m \in \mathbb{C} \label{eq:1rrdm_ac} \\
\edens{e}{u} &= \tens{\alpha}{u}{e} - 2 \sum_{v}^{\mathbb{A}} \tens{\zeta}{e}{v} \density{v}{u}, && \forall\, e \in \mathbb{V}, \forall\, u \in \mathbb{A} \label{eq:1rrdm_va}
\end{align}
where $\cdensity{a}{b} = \kro{a}{b} - \density{a}{b}$ are the 1-hRDM elements
and we also introduce an intermediate $\tens{\alpha}{ij\cdots}{ab\cdots} = 2 \tens{\kappa}{ij\cdots}{ab\cdots} + \tens{\tau}{ab\cdots}{ij\cdots} \, {\cal R}_s (\mpd{ij\cdots}{ab\cdots})$ for convenience.
In deriving Eqs.~\eqref{eq:1rrdm_vv}--\eqref{eq:1rrdm_va}, we have used the fact that $\boldsymbol \zeta$ is symmetric, that is, $\tens{\zeta}{p}{q} = \tens{\zeta}{q}{p}$.

The two-body relaxed density matrix elements can be summarized as follows.
There are seven terms involving $\tens{\alpha}{ij}{ab}$:
\begin{align}
\edens{ef}{mn} &= \frac{1}{4} \tens{\alpha}{mn}{ef}, \\
\edens{ef}{uv} &= \frac{1}{4} \tens{\alpha}{uv}{ef}, \\
\edens{uv}{mn} &= \frac{1}{4} \tens{\alpha}{uv}{ef}, \\
\edens{eu}{mn} &= \frac{1}{2} \tens{\alpha}{mn}{eu}, \\
\edens{ev}{mu} &= \tens{\alpha}{mu}{ev} + \big( \tens{\alpha}{m}{e} + 2 \tens{\zeta}{m}{e} \big) \density{u}{v}, \\
\edens{ey}{ux} &= \frac{1}{2} \tens{\alpha}{ux}{ey} + \tens{\alpha}{u}{e} \density{x}{y} - \sum_{v}^{\mathbb{A}} \tens{\zeta}{v}{e} \density{vy}{ux}, \\
\edens{ux}{my} &= \frac{1}{2} \tens{\alpha}{my}{ux} + \big( \tens{\alpha}{m}{u} + 2 \tens{\zeta}{m}{u} \big) \density{y}{x} - \sum_{v}^{\mathbb{A}} \tens{\zeta}{m}{v} \density{vy}{ux}.
\end{align}
The remaining eight terms are expressed as:
\begin{align}
\edens{em}{fm} &= \edens{e}{f}, \\
\edens{en}{mn} &= \edens{e}{m}, \\
\edens{un}{mn} &= \edens{u}{m}, \\
\edens{em}{um} &= \edens{e}{u}, \\
\edens{mo}{no} &= \frac{1}{2} \kro{m}{n} + \tens{\zeta}{n}{m}, \\
\edens{um}{vn} &= \tens{\zeta}{n}{m} \density{v}{u} + \kro{m}{n} \edens{u}{v}, \\
\edens{eu}{f\!v} &= \tens{\zeta}{f}{e} \density{v}{u}, \\
\edens{uv}{xy} &= \frac{1}{4} \big( \density{uv}{xy} + \mdensity{uv}{xy} \big) + \tens{\zeta}{x}{u} \density{y}{v}.
\end{align}
Here, we follow the index convention in Table \ref{tab:orbital_space} and thus omit the orbital type in the above equations for brevity.

\section{Derivatives in Multiplier Equations}
\label{app:deriv}

\subsection{Modified Integrals and Cluster Amplitudes}
\label{app:kappa_tau}

In this appendix, we derive the expressions for the derivatives given by Eqs.~\eqref{eq:kappa1}, \eqref{eq:kappa2}, \eqref{eq:E2nd_deriv_t1} and \eqref{eq:E2nd_deriv_t2}.
Because $\widetilde{H}$ and $\hat{T}$ include only one- and two-body operators, we only need to derive the fully connected terms from the following three types of commutators $[\no{\sqop{i}{a}}, \no{\sqop{b}{j}}]$, $[\no{\sqop{i}{a}}, \no{\sqop{cd}{kl}}]$, and $[\no{\sqop{ij}{ab}}, \no{\sqop{cd}{kl}}]$.
These commutators are evaluated to be
\begin{align}
[\no{\sqop{i}{a}}, \no{\sqop{b}{j}}]_0 =&\, \density{i}{j} \cdensity{b}{a} - \cdensity{i}{j} \density{b}{a}, \label{eq:comm110} \\
[\no{\sqop{i}{a}}, \no{\sqop{cd}{kl}}]_0 =&\, {\cal P}(cd) \kro{c}{a} \cumulant{id}{kl} - {\cal P}(kl) \kro{i}{k} \cumulant{cd}{al}, \label{eq:comm120} \\
[\no{\sqop{ij}{ab}}, \no{\sqop{cd}{kl}}]_0 =&\, {\cal P}(kl) {\cal P}(cd) ( \density{i}{k} \density{j}{l} \cdensity{c}{a} \cdensity{d}{b} - \cdensity{i}{k} \cdensity{j}{l} \density{c}{a} \density{d}{b} ) \notag\\
&+ {\cal P}(kl) ( \density{i}{k} \density{j}{l} - \cdensity{i}{k} \cdensity{j}{l} ) \cumulant{cd}{ab} \notag\\
&+ {\cal P}(cd) ( \cdensity{c}{a} \cdensity{d}{b} - \density{c}{a} \density{d}{b} ) \cumulant{ij}{kl} \notag\\
&+ {\cal P}(ij) {\cal P}(ab) {\cal P}(kl) {\cal P}(cd) \density{i}{k} \cdensity{c}{a} \cumulant{jd}{bl} \notag\\
&- {\cal P}(ij) {\cal P}(ab) {\cal P}(kl) {\cal P}(cd) \cdensity{i}{k} \density{c}{a} \cumulant{jd}{bl} \notag\\
&+ {\cal P}(ab) {\cal P}(cd) \kro{c}{a} \cumulant{ijd}{kbl} \notag\\
&- {\cal P}(ij) {\cal P}(kl) \kro{i}{k} \cumulant{jcd}{bal}. \label{eq:comm220}
\end{align}
Here, we have introduced the two- and three-body density cumulants defined by:
\begin{align}
\cumulant{xy}{uv} &= \density{xy}{uv} - \density{x}{u} \density{y}{v} + \density{y}{u} \density{x}{v}, \label{eq:2cu} \\
\cumulant{xyz}{uvw} &= \density{xyz}{uvw} - \sum_\pi (-1)^{{\cal N}(\pi)} \density{x}{u} \cumulant{yz}{vw}
- \det(\density{x}{u} \density{y}{v} \density{z}{w}). \label{eq:3cu}
\end{align}
In Eq.~\eqref{eq:3cu}, $\det(\cdot)$ indicates the sum of all permutations of lower (or upper) labels with a sign factor corresponding to the parity of permutations
and $\sum_{\pi} (-1)^{{\cal N}(\pi)}$ indicates a sum over all permutations of the lower and upper labels with a sign factor given by the number of inversions in $\pi$ [${\cal N}(\pi)$].\cite{kutzelnigg1997normal}

It is now easy to check that
\begin{align}
\frac{\partial E^{(2)}}{\partial \tens{\rh}{i}{a}} =&\, \sum_{j}^{\mathbb{H}} \sum_{b}^{\mathbb{P}} \tens{t}{b}{j} \density{i}{j} \cdensity{b}{a} \notag\\
&+ \frac{1}{2} \sum_{kl}^{\mathbb{H}} \sum_{d}^{\mathbb{P}} \tens{t}{ad}{kl} \cumulant{id}{kl}
- \frac{1}{2} \sum_{cd}^{\mathbb{P}} \sum_{l}^{\mathbb{H}} \tens{t}{cd}{il} \cumulant{cd}{al}, \label{eq:E2nd_deriv_kappa1} \\
4\frac{\partial E^{(2)}}{\partial \tens{\rh}{ij}{ab}} =&\, {\cal P}(ab) \sum_{k}^{\mathbb{H}} \tens{t}{a}{k} \cumulant{ij}{kb}
- {\cal P}(ij) \sum_{c}^{\mathbb{P}} \tens{t}{c}{i} \cumulant{cj}{ab} \notag\\
&+ \sum_{kl}^{\mathbb{H}} \sum_{cd}^{\mathbb{P}} \tens{t}{cd}{kl} \big[ \density{i}{k} \density{j}{l} \cdensity{c}{a} \cdensity{d}{b} + {\cal P}(ij) {\cal P}(ab) \density{i}{k} \cdensity{c}{a} \cumulant{jd}{bl} \big] \notag\\
&+ \frac{1}{2} \sum_{kl}^{\mathbb{H}} \sum_{cd}^{\mathbb{P}} \tens{t}{cd}{kl} \big( \density{i}{k} \density{j}{l} \cumulant{cd}{ab} + \cdensity{c}{a} \cdensity{d}{b} \cumulant{ij}{kl} \big) \notag\\
&+ \frac{1}{2} {\cal P}(ab) \sum_{kl}^{\mathbb{H}} \sum_{d}^{\mathbb{P}} \tens{t}{ad}{kl} \cumulant{ijd}{kbl} \notag\\
&- \frac{1}{2} {\cal P}(ij) \sum_{l}^{\mathbb{H}} \sum_{cd}^{\mathbb{P}} \tens{t}{cd}{il} \cumulant{jcd}{bal}. \label{eq:E2nd_deriv_kappa2}
\end{align}
The derivatives of $E^{(2)}$ with respect to amplitudes can be obtained by making the replacements of $\tens{t}{a}{i} \rightarrow \tens{\tilde{h}}{i}{a}$ and $\tens{t}{ab}{ij} \rightarrow \tens{\tilde{h}}{ij}{ab}$ in Eqs.~\eqref{eq:E2nd_deriv_kappa1} and \eqref{eq:E2nd_deriv_kappa2}.

Several properties can be used to further simply Eqs.~\eqref{eq:E2nd_deriv_kappa1} and \eqref{eq:E2nd_deriv_kappa2}.
The 1-pRDMs $\density{p}{q}$ and 1-hRDMs $\cdensity{p}{q}$ possess very simple structures:
\begin{equation}
\density{p}{q} =
\begin{dcases}
\kro{p}{q} & \forall\, p,q \in {\mathbb{C}} \\
\density{p}{q} & \forall\, p,q \in {\mathbb{A}} \\
0 & \text{otherwise}
\end{dcases}
,\quad
\cdensity{p}{q} =
\begin{dcases}
\kro{p}{q} & \forall\, p,q \in {\mathbb{V}} \\
\kro{p}{q} - \density{p}{q} & \forall\, p,q \in {\mathbb{A}} \\
0 & \text{otherwise}
\end{dcases}.
\end{equation}
Density cumulants are only nonzero when all indices are active orbitals, that is, $\cumulant{pq\cdots}{rs\cdots} = 0$ if any of the indices $p,q,r,s,\cdots$ is not active.

\subsection{Orbital Energies}
\label{app:zeta_diag}

For convenience, we first evaluate the derivatives of $\tens{\check{f}}{i}{a}$ and ${\cal R}_s(\mpd{ij\cdots}{ab\cdots})$ with respect to semicanonical orbital energies:
\begin{align}
\pdev{\tens{\check{f}}{i}{a}}{\dfock{p}} &= \sum_{j}^{\mathbb{H}} \sum_{b}^{\mathbb{P}} \tens{t}{ab}{ij} ( \density{p}{j} \kro{b}{p} - \density{b}{p} \kro{p}{j} ), \\
\pdev{{\cal R}_s(\mpd{ij\cdots}{ab\cdots})}{\dfock{p}} &= \big[ 2s \exps{ab\cdots}{ij\cdots} - {\cal R}_s (\mpd{ij\cdots}{ab\cdots}) / \mpd{ij\cdots}{ab\cdots} \big] \tens{\cal D}{ab\cdots}{ij\cdots},
\end{align}
where $\tens{\cal D}{ab\cdots}{ij\cdots} = \pdev*{\mpd{ij\cdots}{ab\cdots}}{\dfock{p}} = \kro{i}{p} + \kro{j}{p} + \cdots - \kro{p}{a} - \kro{p}{b} - \cdots$.
Using these expressions, the partial derivatives in Eq.~\eqref{eq:zeta_diag} are calculated as follows:
\begin{align}
\pdev{ {\cal T}_1 }{\dfock{p}} =&\,
  \sum_{i}^{\mathbb{H}} \sum_{a}^{\mathbb{P}} \tens{\tau}{a}{i}
  \Big[ {\cal R}_s(\mpd{i}{a}) \pdev{\tens{\check{f}}{i}{a}}{\dfock{p}} + \tens{\check{f}}{i}{a} \pdev{{\cal R}_s(\mpd{i}{a})}{\dfock{p}} \Big], \\
\pdev{ {\cal T}_2 }{\dfock{p}} =&\,
  \frac{1}{4} \sum_{ij}^{\mathbb{H}} \sum_{ab}^{\mathbb{P}} \tens{\tau}{ab}{ij} \tens{v}{ij}{ab} \pdev{{\cal R}_s(\mpd{ij}{ab})}{\dfock{p}}, \\
\pdev{ {\cal \tilde{H}}_1 }{\dfock{p}} =&\,
  \sum_{i}^{\mathbb{H}} \sum_{a}^{\mathbb{P}} \tens{\kappa}{i}{a} ( \pdev*{\tens{\check{f}}{i}{a}}{\dfock{p}} ) - \sum_{i}^{\mathbb{H}} \sum_{a}^{\mathbb{P}} \tens{\kappa}{i}{a} \tens{t}{a}{i} \tens{\cal D}{a}{i}, \\
\pdev{ {\cal \tilde{H}}_2 }{\dfock{p}} =&\,
  - \frac{1}{4} \sum_{ij}^{\mathbb{H}} \sum_{ab}^{\mathbb{P}} \tens{\kappa}{ij}{ab} \tens{t}{ab}{ij} \tens{\cal D}{ab}{ij}.
\end{align}

\subsection{Orbital Rotations}
\label{app:zeta}

The derivatives of the Lagrangian amplitude contributions with respect to orbital rotations are given by
\begin{align}
\pdev{{\cal T}_1}{\orbrot{q}{p}} &= \sum_{i}^{\mathbb{H}} \sum_{a}^{\mathbb{P}} \tens{\tilde{\tau}}{a}{i} \pdev{\tens{f}{i}{a}}{\orbrot{q}{p}}, \label{eq:t1_orb} \\
\pdev{{\cal T}_2}{\orbrot{q}{p}} &= \frac{1}{4} \sum_{ij}^{\mathbb{H}} \sum_{ab}^{\mathbb{P}} \tens{\tilde{\tau}}{ab}{ij} \pdev{\tens{v}{ij}{ab}}{\orbrot{q}{p}}, \label{eq:t2_orb}
\end{align}
where $\tens{\tilde{\tau}}{a}{i} = \tens{\tau}{a}{i} \, {\cal R}_s (\mpd{i}{a})$ and $\tens{\tilde{\tau}}{ab}{ij} = \tens{\tau}{ab}{ij} \, {\cal R}_s (\mpd{ij}{ab})$.
The orbital response from modified integrals also appears transparent:
\begin{align}
\pdev{{\cal \tilde{H}}_1}{\orbrot{q}{p}} &= 2 \sum_{i}^{\mathbb{H}} \sum_{a}^{\mathbb{P}} \tens{\kappa}{i}{a} \pdev{\tens{f}{i}{a}}{\orbrot{q}{p}}, \label{eq:htilde1_orb} \\
\pdev{{\cal \tilde{H}}_2}{\orbrot{q}{p}} &= \frac{1}{2} \sum_{ij}^{\mathbb{H}} \sum_{ab}^{\mathbb{P}} \tens{\kappa}{ij}{ab} \pdev{\tens{v}{ij}{ab}}{\orbrot{q}{p}}. \label{eq:htilde2_orb}
\end{align}
The derivatives of the bare integrals with respect to orbital rotations are shown to be:
\begin{align}
\pdev{\tens{f}{r}{s}}{\orbrot{q}{p}} =&\,
\tens{f}{p}{s} \kro{q}{r} + \tens{f}{r}{p} \kro{s}{q} + \indicator{C}{q} \big( \tens{v}{rp}{sq} + \tens{v}{rq}{sp} \big) \notag\\
&+ \indicator{A}{q} \Big[ \sum_{x}^{\mathbb{A}} \big( \tens{v}{rp}{sx} \density{q}{x} + \tens{v}{rx}{sp} \density{x}{q} \big) \Big] , \label{eq:fock_orb} \\
\pdev{\tens{v}{ij}{ab}}{\orbrot{q}{p}} =&\, \tens{v}{pj}{ab} \kro{q}{i} + \tens{v}{ip}{ab} \kro{q}{j} + \tens{v}{ij}{pb} \kro{a}{q} + \tens{v}{ij}{ap} \kro{b}{q}. \label{eq:aptei_orb}
\end{align}
In Eq.~\eqref{eq:fock_orb}, we have introduced the indicator function:
\begin{equation}
\indicator{O}{p} := 
\begin{cases}
1 & \text{if } p \in \mathbb{O}, \\
0 & \text{otherwise.}
\end{cases}
\end{equation}

The orbital response term from the reference energy $E_0$ is well-known from CASSCF orbital conditions:
\begin{equation}
\label{eq:E0_orb}
\pdev{E_0}{\orbrot{q}{p}} = 2 \big[
\indicator{C}{q} \tens{f}{p}{q}
+ \indicator{A}{q} \tens{\tilde{f}}{p}{q} \big] .
\end{equation}
Similar equations can be obtained for the CI term in $\cal L$:
\begin{equation}
\label{eq:Etilde_orb}
\pdev{\cal X}{\orbrot{q}{p}} = 2 \Big[
\indicator{C}{q} \sum_{uv}^{\mathbb{A}} \tens{v}{pu}{qv} \mdensity{u}{v}
+ \indicator{A}{q} \tens{\tilde{f}}{p}{q} (\density{}{} \rightarrow \mdensity{}{}) \Big],
\end{equation}
where $\tens{\tilde{f}}{p}{q} (\density{}{} \rightarrow \mdensity{}{})$ holds a similar form of Eq.~\eqref{eq:fock_tilde} with the pRDMs $\density{}{}$ [Eq.~\eqref{eq:rdm}] replaced by the corresponding mRDMs $\mdensity{}{}$ [Eq.\eqref{eq:mrdm}].

Finally, for the Lagrangian term on orbital constraints ${\cal F}$, we have
\begin{align}
\label{eq:orb_orb}
\pdev{\cal F}{\tens{\vartheta}{p}{q}} 
=\, & \sum_{rs}^{\mathbb{G}} \tens{\zeta}{r}{s} \Big(
\big[ 1 - \overline{\indicator{A}{r} \indicator{V}{s}} \big] \pdev{\tens{f}{r}{s}}{\tens{\vartheta}{p}{q}}
- \overline{\indicator{A}{r} \indicator{D}{s}} \pdev{\tens{\tilde{f}}{r}{s}}{\tens{\vartheta}{p}{q}}
\Big) ,
\end{align}
where we adopt the short-hand notation $\overline{\indicator{A}{r} \indicator{V}{s}} = \indicator{A}{r} \indicator{V}{s} + \indicator{A}{s} \indicator{V}{r}$ and $\mathbb{D} \equiv \mathbb{C} \cup \mathbb{V}$.
The derivatives $\pdev*{\tens{\tilde{f}}{r}{u}}{\tens{\vartheta}{p}{q}}$ in Eq.~\eqref{eq:orb_orb} are worked out to be
\begin{align}
\pdev{\tens{\tilde{f}}{r}{u}}{\tens{\vartheta}{p}{q}} =&\,
\tens{\tilde{f}}{p}{u} \kro{q}{r}
+ \indicator{C}{q} \Big[ \sum_{v}^{\mathbb{A}} ( \tens{v}{rp}{vq} + \tens{v}{rq}{vp} ) \density{u}{v} \Big] \notag\\
&+ \indicator{A}{q} \Big[ \tens{\bar{f}}{r}{p} \density{u}{q} + \sum_{xy}^{\mathbb{A}} \Big( \frac{1}{2} \tens{v}{rp}{xy} \density{uq}{xy} + \tens{v}{rx}{py} \density{ux}{qy} \Big) \Big] .
\label{eq:ftilde_orb}
\end{align}

\subsection{CI Coefficients}
\label{app:xi}

We now evaluate all terms in Eq.~\eqref{eq:xi}.
It is straightforward to see that the $\pdev*{\cal X}{c_I}$ term yields:
\begin{equation}
\label{eq:cixdev}
\pdev{\cal X}{c_I} = \sum_{J}^{{\cal M}_0} \xi_J \braket{\Phi_J | \hat{H}^{\rm a} | \Phi_I},
\end{equation}
where the active part of the bare Hamiltonian is defined by
\begin{equation}
\hat{H}^{\rm a} = \sum_{uv}^{\mathbb{A}} \tens{\bar{f}}{u}{v} \sqop{u}{v} + \frac{1}{4} \sum_{uvxy}^{\mathbb{A}} \tens{v}{uv}{xy} \sqop{uv}{xy}
\end{equation}
The derivatives of ${\cal E} + {\cal T}_1 + {\cal \tilde{H}}_1 + {\cal F}$ with respect to $c_I$ can generally be written as
\begin{align}
\label{eq:Gci}
\pdev{}{c_I} ({\cal E} + {\cal T}_1 + {\cal \tilde{H}}_1 + {\cal F}) =&\,
\sum_{uv}^{\mathbb{A}} \tens{g}{v}{u} \pdev{\density{v}{u}}{c_I}
+ \sum_{uvxy}^{\mathbb{A}} \tens{g}{xy}{uv} \pdev{\density{xy}{uv}}{c_I} \notag\\
&+ \sum_{uvwxyz}^{\mathbb{A}} \tens{g}{xyz}{uvw} \pdev{\density{xyz}{uvw}}{c_I} .
\end{align}

Here, the effective integrals are given by
\begin{align}
\tens{g}{v}{u} =&\,
\tens{\bar{f}}{v}{u}
+ \pdev{E^{(2)}}{\density{v}{u}}
+ \sum_{i}^{\mathbb{H}} \sum_{a}^{\mathbb{P}} \big[
\tens{\alpha}{i}{a} \tens{v}{iv}{au} + ( \tens{\tilde{\tau}}{a}{i} + \tens{\kappa}{i}{a} ) \mpd{v}{u} \tens{t}{av}{iu}
\big] \notag\\
&+ \sum_{rs}^{\mathbb{G}} \big[ 1 - \overline{\indicator{A}{r} \indicator{V}{s}} \big] \tens{\zeta}{r}{s} \tens{v}{rv}{su}
- 2 \sum_{r}^{\mathbb{D}} \tens{\zeta}{r}{u} \tens{\bar{f}}{v}{r}
, \label{eq:dci_g1} \\
\tens{g}{xy}{uv} =&\,
\frac{1}{4} \tens{v}{xy}{uv}
+ \pdev{E^{(2)}}{\density{xy}{uv}}
- \sum_{r}^{\mathbb{D}} \tens{\zeta}{r}{u} \tens{v}{xy}{rv}, \label{eq:dci_g2} \\
\tens{g}{xyz}{uvw} =&\,
\frac{1}{4} \sum_{m}^{\mathbb{C}} \tens{\rh}{mz}{uv} \tens{t}{xy}{mw}
- \frac{1}{4} \sum_{e}^{\mathbb{V}} \tens{\rh}{xy}{ew} \tens{t}{ez}{uv}. \label{eq:dci_g3}
\end{align}
In Eqs.~\eqref{eq:dci_g1} and \eqref{eq:dci_g2}, the partial derivatives of $E^{(2)}$ with respect to $\density{v}{u}$ and $\density{xy}{uv}$ are given by:
\begin{align}
\pdev{E^{(2)}}{\density{v}{u}} =&\,
+ \sum_{e}^{\mathbb{V}} \tens{\rh}{v}{e} \tens{t}{e}{u} - \sum_{m}^{\mathbb{C}} \tens{\rh}{m}{u} \tens{t}{v}{m} \notag\\
&+ \frac{1}{2} \sum_{ab}^{\mathbb{P}} \sum_{m}^{\mathbb{C}} \tens{\rh}{mv}{ab} \tens{t}{ab}{mu} - \frac{1}{2} \sum_{a}^{\mathbb{P}} \sum_{mn}^{\mathbb{C}} \tens{\rh}{mn}{au} \tens{t}{av}{mn} \notag\\
&+ \sum_{xy}^{\mathbb{A}} \density{y}{x} \Big[ {\cal P}(ux) \sum_{m}^{\mathbb{C}} \tens{\rh}{m}{x} \tens{t}{vy}{um} - {\cal P}(vy) \sum_{e}^{\mathbb{V}} \tens{\rh}{v}{e} \tens{t}{ey}{ux} \Big] \notag\\
&+ \sum_{xy}^{\mathbb{A}} \density{y}{x} \Big[ {\cal P}(vy) \sum_{m}^{\mathbb{C}} \tens{\tilde{v}}{my}{ux} \tens{t}{v}{m} - {\cal P}(ux) \sum_{e}^{\mathbb{V}} \tens{\rh}{vy}{ex} \tens{t}{e}{u} \Big] \notag\\
&- \sum_{xy}^{\mathbb{A}} \density{y}{x} \sum_{a}^{\mathbb{P}} \sum_{m}^{\mathbb{C}} ( \tens{\rh}{mv}{au} \tens{t}{ay}{mx} + \tens{\rh}{my}{ax} \tens{t}{av}{mu} ) \notag\\
&+ \frac{1}{2} \sum_{wxyz}^{\mathbb{A}} \cumulant{yz}{wx} \sum_{e}^{\mathbb{V}} ( \tens{\rh}{yv}{eu} \tens{t}{ez}{xw} + \tens{\rh}{yz}{ex} \tens{t}{ev}{uw} ) \notag\\
&- \frac{1}{2} \sum_{wxyz}^{\mathbb{A}} \cumulant{yz}{wx} \sum_{m}^{\mathbb{C}} ( \tens{\rh}{mz}{xw} \tens{t}{yv}{mu} + \tens{\rh}{mv}{uw} \tens{t}{yz}{mx} ) \notag\\
&+ \sum_{wxyz}^{\mathbb{A}} \density{z}{y} \density{x}{w} \Big( \sum_{m}^{\mathbb{C}} \tens{\rh}{mz}{uy} \tens{t}{xv}{mw} - \sum_{e}^{\mathbb{V}} \tens{\rh}{xv}{ew} \tens{t}{ez}{uy} \Big) ,
\\
\pdev{E^{(2)}}{\density{xy}{uv}} =&\,
+ \frac{1}{2} \sum_{e}^{\mathbb{V}} \tens{\rh}{x}{e} \tens{t}{ey}{uv} - \frac{1}{2} \sum_{m}^{\mathbb{C}} \tens{\rh}{m}{v} \tens{t}{xy}{um} \notag\\
&+ \frac{1}{2} \sum_{e}^{\mathbb{V}} \tens{\rh}{xy}{ev} \tens{t}{e}{u} - \frac{1}{2} \sum_{m}^{\mathbb{C}} \tens{\tilde{v}}{my}{uv} \tens{t}{x}{m} \notag\\
&+ \frac{1}{8} \sum_{ab}^{\mathbb{P}} \tens{\rh}{xy}{ab} \tens{t}{ab}{uv} + \frac{1}{8} \sum_{mn}^{\mathbb{C}} \tens{\rh}{mn}{uv} \tens{t}{xy}{mn} + \sum_{a}^{\mathbb{P}} \sum_{m}^{\mathbb{C}} \tens{\rh}{mx}{au} \tens{t}{ay}{mv} \notag\\
&+ \frac{1}{2} \sum_{wz}^{\mathbb{A}} \density{z}{w} \sum_{e}^{\mathbb{V}} ( \tens{\rh}{yz}{ew} \tens{t}{ex}{uv} + \tens{\rh}{xy}{eu} \tens{t}{ez}{vw} ) \notag\\
&- \frac{1}{2} \sum_{wz}^{\mathbb{A}} \density{z}{w} \sum_{m}^{\mathbb{C}} ( \tens{\rh}{mx}{uv} \tens{t}{yz}{mw} + \tens{\rh}{mz}{vw} \tens{t}{xy}{mu} ).
\end{align}
The RDM derivatives in Eq.~\eqref{eq:Gci} are generically written as
\begin{equation}
\pdev{\density{uv\dots}{xy\dots}}{c_I} = \braket{\Phi_I | \sqop{uv\dots}{xy\dots} + \sqop{xy\dots}{uv\dots} | \mref }.
\end{equation}
As such, the one- and two-body terms of Eq.~\eqref{eq:Gci} can be computed using a standard CI sigma build with the revised integrals $\tens{g}{v}{u}$ and $\tens{g}{xy}{uv}$.
In this work, we store the derivatives $\pdev*{\density{xyz}{uvw}}{c_I}$ appeared in Eq.~\eqref{eq:Gci}, which requires further optimizations in the future.

Finally, for the $\cal Y$ term, we have
\begin{align}
\pdev*{\cal Y}{c_I} = - 2 \iota c_I.
\end{align}
The multiplier $\iota$ can be easily obtained from Eq.~\eqref{eq:iota} using Eq.~\eqref{eq:Gci}:
\begin{equation}
\iota = \sum_{uv}^{\mathbb{A}} \tens{g}{v}{u} \density{v}{u}
+ \sum_{uvxy}^{\mathbb{A}} \tens{g}{xy}{uv} \density{xy}{uv}
+ \sum_{uvwxyz}^{\mathbb{A}} \tens{g}{xyz}{uvw} \density{xyz}{uvw},
\end{equation}
using the fact that $\density{uv\cdots}{xy\cdots} = \frac{1}{2} \sum_{I}^{{\cal M}_0} c_I \pdev*{\density{uv\cdots}{xy\cdots}}{c_I}$.
Note that $\iota$ depends on the orbital multipliers $\tens{\zeta}{p}{q}$ and changes every iteration of solving the linear system [Eq.~\eqref{eq:linearaxb}].

\section{The Response Equation for the Orbital and CI Coefficients}
\label{app:cpcasscf}

We are now equipped to show all the components of the coupled linear system [Eq.~\eqref{eq:linearaxb}].
The four blocks of the coefficient matrix $\bf A$ can be written in the partial derivative form:
\begin{align}
A^{\rm oo}_{pq, rs} &= \pdev{}{\tens{\zeta}{r}{s}} {\cal P}(pq) \Big( \pdev{\cal F}{\tens{\vartheta}{p}{q}} \Big) = \pdev{\tens{F}{r}{s}}{\tens{\vartheta}{p}{q}} - \pdev{\tens{F}{r}{s}}{\tens{\vartheta}{q}{p}} , \label{eq:Aoo_dev} \\
A^{\rm oc}_{pq, J} &= \pdev{}{\xi_J} {\cal P}(pq) \Big( \pdev{\cal X}{\tens{\vartheta}{p}{q}} \Big) = \pdev{X_J}{\tens{\vartheta}{p}{q}} - \pdev{X_J}{\tens{\vartheta}{q}{p}} , \label{eq:Aoc_dev} \\
A^{\rm co}_{I, rs} &= \pdev{}{\tens{\zeta}{r}{s}} \pdev{}{c_I} {\cal F} = \pdev{\tens{F}{r}{s}}{c_I} , \label{eq:Aco_dev} \\
A^{\rm cc}_{I,J} &= \pdev{}{\xi_J} \pdev{}{c_I} {\cal X} = \pdev{X_J}{c_I}. \label{eq:Acc_dev}
\end{align}
As such, the orbital [Eq.~\eqref{eq:orb_response}] and CI [Eq.~\eqref{eq:ci_response}] response equations can be written as
\begin{align}
\sum_{rs}^{\mathbb{G}} A^{\rm oo}_{pq, rs} \tens{\zeta}{r}{s} + \sum_{J}^{{\cal M}_0} A^{\rm oc}_{pq, J} \xi_J &= b^{\rm o}_{pq}, \label{eq:orb_response_element} \\
\sum_{rs}^{\mathbb{G}} A^{\rm co}_{I, rs} \tens{\zeta}{r}{s} + \sum_{J}^{{\cal M}_0} A^{\rm cc}_{I, J} \xi_J &= b^{\rm c}_{I}. \label{eq:ci_response_element}
\end{align}
The block elements of the $\bf b$ vector on the right-hand-side of Eqs.~\eqref{eq:orb_response_element} and \eqref{eq:ci_response_element} are given by
\begin{align}
b^{\rm o}_{pq} &= - {\cal P}(pq) \Big( \pdev{}{\tens{\vartheta}{p}{q}} \big[ E_0 + \sum_{n=1}^{2} ( {\cal T}_n + {\cal H}_n ) \big] \Big), \label{eq:bo_pq} \\
b^{\rm c}_{I} &= - \pdev{}{c_I} ( {\cal E} +  {\cal T}_1 + {\cal H}_1 + {\cal Y}). \label{eq:bc_I}
\end{align}
For $b^{\rm o}_{pq}$, all components of Eq.~\eqref{eq:bo_pq} have been reported in Sec.~\ref{app:zeta}, specifically Eqs.~\eqref{eq:t1_orb}--\eqref{eq:E0_orb}.
The expression of $b^{\rm c}_{I}$ can be obtained using Eqs.~\eqref{eq:Gci}--\eqref{eq:dci_g3} by omitting the $\tens{\zeta}{r}{s}$ contributions in Eqs.~\eqref{eq:dci_g1} and \eqref{eq:dci_g2}.

We may further express $A^{\rm oo}_{pq, rs}$ [Eq.~\eqref{eq:Aoo_dev}] in terms of $\tens{f}{r}{s}$ and $\tens{\tilde{f}}{r}{s}$, resulting in the following cases:
\begin{align}
\label{eq:Aoo_pqrs}
A^{\rm oo}_{pq, rs} =&\,
{\cal P}(pq) \big( \big[ 1 - \overline{\indicator{V}{r} \indicator{A}{s}} \big] ( \pdev*{\tens{f}{r}{s}}{\tens{\vartheta}{p}{q}} ) \big) \notag\\
&- {\cal P}(pq) \big[ \overline{\indicator{D}{r} \indicator{A}{s}} ( \pdev*{\tens{\tilde{f}}{r}{s}}{\tens{\vartheta}{p}{q}} ) \big] .
\end{align}
The partial derivatives appeared in Eq.~\eqref{eq:Aoo_pqrs} are reported in Eqs.~\eqref{eq:fock_orb} and \eqref{eq:ftilde_orb}.
Simplifications may be achieved by utilizing the CASSCF semicanonical orbital constraint [Eqs.~\eqref{eq:casscf_fock} and \eqref{eq:fcons}].
For example, when all $p, q, r, s \in \mathbb{C}$, the expression of $A^{\rm oo}_{pq, rs}$ is simply
\begin{equation}
A^{\rm oo}_{pq, rs} = \tens{\Delta}{q}{p} \big( \kro{q}{r} \kro{s}{p} + \kro{p}{r} \kro{s}{q} \big), \quad \forall\, p, q, r, s \in \mathbb{C}. 
\end{equation}
In Eq.~\eqref{eq:Aoc_dev}, the partial derivatives $\pdev*{X_J}{\tens{\vartheta}{p}{q}}$ yield:
\begin{align}
\Big( \pdev{X_J}{\tens{\vartheta}{p}{q}} \Big)^\perp =&\,
\indicator{A}{q} \Big( 2 \sum_{v}^{\mathbb{A}} \tens{\bar{f}}{p}{v} \pdev{\mdensity{q}{v}}{\xi_J} + \sum_{vxy}^{\mathbb{A}} \tens{v}{pv}{xy} \pdev{\mdensity{qv}{xy}}{\xi_J} \Big) \notag\\
&+ \indicator{C}{q} \Big( 2 \sum_{uv}^{\mathbb{A}} \tens{v}{pu}{qv} \pdev{\mdensity{u}{v}}{\xi_J} \Big) ,
\end{align}
where $\pdev*{\mdensity{uv\cdots}{xy\cdots}}{\xi_J} = \braket{\Phi_J | \sqop{uv\cdots}{xy\cdots} | \mref}$ and we only keep those components that are perpendicular to the CI vector $\bf c$.
For the $A^{\rm co}_{I, rs}$ term [Eq.~\eqref{eq:Aco_dev}], we write
\begin{equation}
A^{\rm co}_{I, rs} =
\big[ 1 - \overline{\indicator{V}{r} \indicator{A}{s}} \big] \Big( \pdev{\tens{f}{r}{s}}{c_I} \Big)
- \overline{\indicator{D}{r} \indicator{A}{s}} \Big( \pdev{\tens{\tilde{f}}{r}{s}}{c_I} \Big) .
\end{equation}
Here, the derivatives of $\tens{f}{r}{s}$ and $\tens{\tilde{f}}{r}{s}$ with respect to $c_I$ are evaluated to be
\begin{align}
\pdev{\tens{f}{r}{s}}{c_I} &= \sum_{uv}^{\mathbb{A}} \tens{v}{rv}{su} \pdev{\density{v}{u}}{c_I}, \\
\pdev{\tens{\tilde{f}}{r}{s}}{c_I} &= \sum_{v}^{\mathbb{A}} \tens{\bar{f}}{p}{v} \pdev{\density{s}{v}}{c_I} + \frac{1}{2} \sum_{vxy}^{\mathbb{A}} \tens{v}{pv}{xy} \pdev{\density{qv}{xy}}{c_I} .
\end{align}
Lastly, the $A^{\rm cc}_{I,J}$ term [Eq.~\eqref{eq:Acc_dev}] can be easily derived using Eq.~\eqref{eq:cixdev}:
\begin{equation}
A^{\rm cc}_{I,J} = \braket{\Phi_I | \hat{H}^{\rm a} | \Phi_J}.
\end{equation}

\titleformat{\section}
{\normalfont\large\sffamily\bfseries\color{goodred}}
{\raisebox{5pt}{\colorbox{goodred}{\rule[-2pt]{2pt}{2pt}}}}{0.5em}{\uppercase}

%

\section{Acknowledgements}
The authors would like to acknowledge helpful discussions with Toru Shiozaki, Yuanzhe Xi, Haruyuki Nakano, and Alexander Sokolov.
This work was supported by the U.S. Department of Energy under Award No.~DE-SC0016004.
F.A.E. acknowledges support from a Camille Dreyfus Teacher-Scholar Award.

\bibliography{mrgradient_reference,bib-ext}

\providecommand{\latin}[1]{#1}
\makeatletter
\providecommand{\doi}
  {\begingroup\let\do\@makeother\dospecials
  \catcode`\{=1 \catcode`\}=2 \doi@aux}
\providecommand{\doi@aux}[1]{\endgroup\texttt{#1}}
\makeatother
\providecommand*\mcitethebibliography{\thebibliography}
\csname @ifundefined\endcsname{endmcitethebibliography}
  {\let\endmcitethebibliography\endthebibliography}{}
\begin{mcitethebibliography}{96}
\providecommand*\natexlab[1]{#1}
\providecommand*\mciteSetBstSublistMode[1]{}
\providecommand*\mciteSetBstMaxWidthForm[2]{}
\providecommand*\mciteBstWouldAddEndPuncttrue
  {\def\EndOfBibitem{\unskip.}}
\providecommand*\mciteBstWouldAddEndPunctfalse
  {\let\EndOfBibitem\relax}
\providecommand*\mciteSetBstMidEndSepPunct[3]{}
\providecommand*\mciteSetBstSublistLabelBeginEnd[3]{}
\providecommand*\EndOfBibitem{}
\mciteSetBstSublistMode{f}
\mciteSetBstMaxWidthForm{subitem}{(\alph{mcitesubitemcount})}
\mciteSetBstSublistLabelBeginEnd
  {\mcitemaxwidthsubitemform\space}
  {\relax}
  {\relax}

\bibitem[Abbott \latin{et~al.}(2021)Abbott, Abbott, Turney, and
  Schaefer]{abbott2021arbitrary}
Abbott,~A.~S.; Abbott,~B.~Z.; Turney,~J.~M.; Schaefer,~H.~F. \emph{J. Phys.
  Chem. Lett.} \textbf{2021}, \emph{12}, 3232--3239\relax
\mciteBstWouldAddEndPuncttrue
\mciteSetBstMidEndSepPunct{\mcitedefaultmidpunct}
{\mcitedefaultendpunct}{\mcitedefaultseppunct}\relax
\EndOfBibitem
\bibitem[Iftimie \latin{et~al.}(2005)Iftimie, Minary, and
  Tuckerman]{Iftimie.2005}
Iftimie,~R.; Minary,~P.; Tuckerman,~M.~E. \emph{Proc. Natl. Acad. Sci. U.S.A.}
  \textbf{2005}, \emph{102}, 6654--6659\relax
\mciteBstWouldAddEndPuncttrue
\mciteSetBstMidEndSepPunct{\mcitedefaultmidpunct}
{\mcitedefaultendpunct}{\mcitedefaultseppunct}\relax
\EndOfBibitem
\bibitem[Park and Shiozaki(2017)Park, and Shiozaki]{Park:2017jb}
Park,~J.~W.; Shiozaki,~T. \emph{J. Chem. Theory Comput.} \textbf{2017},
  \emph{13}, 3676--3683\relax
\mciteBstWouldAddEndPuncttrue
\mciteSetBstMidEndSepPunct{\mcitedefaultmidpunct}
{\mcitedefaultendpunct}{\mcitedefaultseppunct}\relax
\EndOfBibitem
\bibitem[Curchod and Martínez(2018)Curchod, and Martínez]{Curchod.2018}
Curchod,~B. F.~E.; Martínez,~T.~J. \emph{Chem. Rev.} \textbf{2018},
  \emph{118}, 3305--3336\relax
\mciteBstWouldAddEndPuncttrue
\mciteSetBstMidEndSepPunct{\mcitedefaultmidpunct}
{\mcitedefaultendpunct}{\mcitedefaultseppunct}\relax
\EndOfBibitem
\bibitem[Pinski and Neese(2018)Pinski, and Neese]{Pinski2018}
Pinski,~P.; Neese,~F. \emph{J. Chem. Phys.} \textbf{2018}, \emph{148},
  31101\relax
\mciteBstWouldAddEndPuncttrue
\mciteSetBstMidEndSepPunct{\mcitedefaultmidpunct}
{\mcitedefaultendpunct}{\mcitedefaultseppunct}\relax
\EndOfBibitem
\bibitem[Dornbach and Werner(2019)Dornbach, and Werner]{Dornbach2019}
Dornbach,~M.; Werner,~H.-J. \emph{Mol. Phys.} \textbf{2019}, \emph{117},
  1252--1263\relax
\mciteBstWouldAddEndPuncttrue
\mciteSetBstMidEndSepPunct{\mcitedefaultmidpunct}
{\mcitedefaultendpunct}{\mcitedefaultseppunct}\relax
\EndOfBibitem
\bibitem[Ni \latin{et~al.}(2019)Ni, Wang, Li, Pulay, and Li]{Ni:2019cd}
Ni,~Z.; Wang,~Y.; Li,~W.; Pulay,~P.; Li,~S. \emph{J. Chem. Theory Comput.}
  \textbf{2019}, \emph{15}, 3623--3634\relax
\mciteBstWouldAddEndPuncttrue
\mciteSetBstMidEndSepPunct{\mcitedefaultmidpunct}
{\mcitedefaultendpunct}{\mcitedefaultseppunct}\relax
\EndOfBibitem
\bibitem[Szalay \latin{et~al.}(2012)Szalay, M{\"u}ller, Gidofalvi, Lischka, and
  Shepard]{Szalay:2012df}
Szalay,~P.~G.; M{\"u}ller,~T.; Gidofalvi,~G.; Lischka,~H.; Shepard,~R.
  \emph{Chem. Rev.} \textbf{2012}, \emph{112}, 108--181\relax
\mciteBstWouldAddEndPuncttrue
\mciteSetBstMidEndSepPunct{\mcitedefaultmidpunct}
{\mcitedefaultendpunct}{\mcitedefaultseppunct}\relax
\EndOfBibitem
\bibitem[Lyakh \latin{et~al.}(2012)Lyakh, Musia{\l}, Lotrich, and
  Bartlett]{Lyakh:2012cn}
Lyakh,~D.~I.; Musia{\l},~M.; Lotrich,~V.~F.; Bartlett,~R.~J. \emph{Chem. Rev.}
  \textbf{2012}, \emph{112}, 182--243\relax
\mciteBstWouldAddEndPuncttrue
\mciteSetBstMidEndSepPunct{\mcitedefaultmidpunct}
{\mcitedefaultendpunct}{\mcitedefaultseppunct}\relax
\EndOfBibitem
\bibitem[K{\"o}hn \latin{et~al.}(2013)K{\"o}hn, Hanauer, M{\"u}ck, Jagau, and
  Gauss]{Koehn:2013cp}
K{\"o}hn,~A.; Hanauer,~M.; M{\"u}ck,~L.~A.; Jagau,~T.-C.; Gauss,~J. \emph{Wiley
  Interdiscip. Rev. Comput. Mol. Sci.} \textbf{2013}, \emph{3}, 176--197\relax
\mciteBstWouldAddEndPuncttrue
\mciteSetBstMidEndSepPunct{\mcitedefaultmidpunct}
{\mcitedefaultendpunct}{\mcitedefaultseppunct}\relax
\EndOfBibitem
\bibitem[Evangelista(2018)]{Evangelista:2018bt}
Evangelista,~F.~A. \emph{J. Chem. Phys.} \textbf{2018}, \emph{149},
  030901\relax
\mciteBstWouldAddEndPuncttrue
\mciteSetBstMidEndSepPunct{\mcitedefaultmidpunct}
{\mcitedefaultendpunct}{\mcitedefaultseppunct}\relax
\EndOfBibitem
\bibitem[Andersson \latin{et~al.}(1990)Andersson, Malmqvist, Roos, Sadlej, and
  Wolinski]{andersson1990second}
Andersson,~K.; Malmqvist,~P.~A.; Roos,~B.~O.; Sadlej,~A.~J.; Wolinski,~K.
  \emph{J. Phys. Chem.} \textbf{1990}, \emph{94}, 5483--5488\relax
\mciteBstWouldAddEndPuncttrue
\mciteSetBstMidEndSepPunct{\mcitedefaultmidpunct}
{\mcitedefaultendpunct}{\mcitedefaultseppunct}\relax
\EndOfBibitem
\bibitem[Hirao(1992)]{hirao1992multireference}
Hirao,~K. \emph{Chem. Phys. Lett.} \textbf{1992}, \emph{190}, 374--380\relax
\mciteBstWouldAddEndPuncttrue
\mciteSetBstMidEndSepPunct{\mcitedefaultmidpunct}
{\mcitedefaultendpunct}{\mcitedefaultseppunct}\relax
\EndOfBibitem
\bibitem[Andersson \latin{et~al.}(1992)Andersson, Malmqvist, and
  Roos]{andersson1992second}
Andersson,~K.; Malmqvist,~P.-{\AA}.; Roos,~B.~O. \emph{J. Chem. Phys.}
  \textbf{1992}, \emph{96}, 1218--1226\relax
\mciteBstWouldAddEndPuncttrue
\mciteSetBstMidEndSepPunct{\mcitedefaultmidpunct}
{\mcitedefaultendpunct}{\mcitedefaultseppunct}\relax
\EndOfBibitem
\bibitem[Kozlowski and Davidson(1994)Kozlowski, and Davidson]{Kozlowski:1994cw}
Kozlowski,~P.~M.; Davidson,~E.~R. \emph{J. Chem. Phys.} \textbf{1994},
  \emph{100}, 3672--3682\relax
\mciteBstWouldAddEndPuncttrue
\mciteSetBstMidEndSepPunct{\mcitedefaultmidpunct}
{\mcitedefaultendpunct}{\mcitedefaultseppunct}\relax
\EndOfBibitem
\bibitem[Werner(1996)]{werner1996third}
Werner,~H.-J. \emph{Mol. Phys.} \textbf{1996}, \emph{89}, 645--661\relax
\mciteBstWouldAddEndPuncttrue
\mciteSetBstMidEndSepPunct{\mcitedefaultmidpunct}
{\mcitedefaultendpunct}{\mcitedefaultseppunct}\relax
\EndOfBibitem
\bibitem[Mahapatra \latin{et~al.}(1999)Mahapatra, Datta, and
  Mukherjee]{Mahapatra:1999gh}
Mahapatra,~U.~S.; Datta,~B.; Mukherjee,~D. \emph{Chem. Phys. Lett.}
  \textbf{1999}, \emph{299}, 42--50\relax
\mciteBstWouldAddEndPuncttrue
\mciteSetBstMidEndSepPunct{\mcitedefaultmidpunct}
{\mcitedefaultendpunct}{\mcitedefaultseppunct}\relax
\EndOfBibitem
\bibitem[Angeli \latin{et~al.}(2001)Angeli, Cimiraglia, Evangelisti, Leininger,
  and Malrieu]{angeli2001introduction}
Angeli,~C.; Cimiraglia,~R.; Evangelisti,~S.; Leininger,~T.; Malrieu,~J.-P.
  \emph{J. Chem. Phys.} \textbf{2001}, \emph{114}, 10252--10264\relax
\mciteBstWouldAddEndPuncttrue
\mciteSetBstMidEndSepPunct{\mcitedefaultmidpunct}
{\mcitedefaultendpunct}{\mcitedefaultseppunct}\relax
\EndOfBibitem
\bibitem[Angeli \latin{et~al.}(2002)Angeli, Cimiraglia, and
  Malrieu]{angeli2002n}
Angeli,~C.; Cimiraglia,~R.; Malrieu,~J.-P. \emph{J. Chem. Phys.} \textbf{2002},
  \emph{117}, 9138--9153\relax
\mciteBstWouldAddEndPuncttrue
\mciteSetBstMidEndSepPunct{\mcitedefaultmidpunct}
{\mcitedefaultendpunct}{\mcitedefaultseppunct}\relax
\EndOfBibitem
\bibitem[Khait \latin{et~al.}(2002)Khait, Song, and Hoffmann]{Khait:2002ca}
Khait,~Y.~G.; Song,~J.; Hoffmann,~M.~R. \emph{J. Chem. Phys.} \textbf{2002},
  \emph{117}, 4133--4145\relax
\mciteBstWouldAddEndPuncttrue
\mciteSetBstMidEndSepPunct{\mcitedefaultmidpunct}
{\mcitedefaultendpunct}{\mcitedefaultseppunct}\relax
\EndOfBibitem
\bibitem[Szabados \latin{et~al.}(2005)Szabados, Rolik, T{\'o}th, and
  Surj{\'a}n]{Szabados:2005jk}
Szabados,~{\'A}.; Rolik,~Z.; T{\'o}th,~G.; Surj{\'a}n,~P.~R. \emph{J. Chem.
  Phys.} \textbf{2005}, \emph{122}, 114104\relax
\mciteBstWouldAddEndPuncttrue
\mciteSetBstMidEndSepPunct{\mcitedefaultmidpunct}
{\mcitedefaultendpunct}{\mcitedefaultseppunct}\relax
\EndOfBibitem
\bibitem[Chaudhuri \latin{et~al.}(2005)Chaudhuri, Freed, Hose, Piecuch,
  Kowalski, W{\l}och, Chattopadhyay, Mukherjee, Rolik, Szabados, \latin{et~al.}
  others]{chaudhuri2005comparison}
Chaudhuri,~R.~K.; Freed,~K.~F.; Hose,~G.; Piecuch,~P.; Kowalski,~K.;
  W{\l}och,~M.; Chattopadhyay,~S.; Mukherjee,~D.; Rolik,~Z.; Szabados,~{\'A}.,
  \latin{et~al.}  \emph{J. Chem. Phys.} \textbf{2005}, \emph{122}, 134105\relax
\mciteBstWouldAddEndPuncttrue
\mciteSetBstMidEndSepPunct{\mcitedefaultmidpunct}
{\mcitedefaultendpunct}{\mcitedefaultseppunct}\relax
\EndOfBibitem
\bibitem[Hoffmann \latin{et~al.}(2009)Hoffmann, Datta, Das, Mukherjee,
  Szabados, Rolik, and Surj{\'a}n]{hoffmann2009comparative}
Hoffmann,~M.~R.; Datta,~D.; Das,~S.; Mukherjee,~D.; Szabados,~A.; Rolik,~Z.;
  Surj{\'a}n,~P.~R. \emph{J. Chem. Phys.} \textbf{2009}, \emph{131},
  204104\relax
\mciteBstWouldAddEndPuncttrue
\mciteSetBstMidEndSepPunct{\mcitedefaultmidpunct}
{\mcitedefaultendpunct}{\mcitedefaultseppunct}\relax
\EndOfBibitem
\bibitem[Evangelista \latin{et~al.}(2009)Evangelista, Simmonett, Schaefer,
  Mukherjee, and Allen]{evangelista2009companion}
Evangelista,~F.~A.; Simmonett,~A.~C.; Schaefer,~H.~F.; Mukherjee,~D.;
  Allen,~W.~D. \emph{Phys. Chem. Chem. Phys.} \textbf{2009}, \emph{11},
  4728--4741\relax
\mciteBstWouldAddEndPuncttrue
\mciteSetBstMidEndSepPunct{\mcitedefaultmidpunct}
{\mcitedefaultendpunct}{\mcitedefaultseppunct}\relax
\EndOfBibitem
\bibitem[Sokolov \latin{et~al.}(2017)Sokolov, Guo, Ronca, and
  Chan]{Sokolov:2017gr}
Sokolov,~A.~Y.; Guo,~S.; Ronca,~E.; Chan,~G. K.-L. \emph{J. Chem. Phys.}
  \textbf{2017}, \emph{146}, 244102\relax
\mciteBstWouldAddEndPuncttrue
\mciteSetBstMidEndSepPunct{\mcitedefaultmidpunct}
{\mcitedefaultendpunct}{\mcitedefaultseppunct}\relax
\EndOfBibitem
\bibitem[Giner \latin{et~al.}(2017)Giner, Angeli, Garniron, Scemama, and
  Malrieu]{giner2017jeziorski}
Giner,~E.; Angeli,~C.; Garniron,~Y.; Scemama,~A.; Malrieu,~J.-P. \emph{J. Chem.
  Phys.} \textbf{2017}, \emph{146}, 224108\relax
\mciteBstWouldAddEndPuncttrue
\mciteSetBstMidEndSepPunct{\mcitedefaultmidpunct}
{\mcitedefaultendpunct}{\mcitedefaultseppunct}\relax
\EndOfBibitem
\bibitem[Roos and Andersson(1995)Roos, and
  Andersson]{roos1995multiconfigurational}
Roos,~B.~O.; Andersson,~K. \emph{Chem. Phys. Lett.} \textbf{1995}, \emph{245},
  215--223\relax
\mciteBstWouldAddEndPuncttrue
\mciteSetBstMidEndSepPunct{\mcitedefaultmidpunct}
{\mcitedefaultendpunct}{\mcitedefaultseppunct}\relax
\EndOfBibitem
\bibitem[Forsberg and Malmqvist(1997)Forsberg, and Malmqvist]{Forsberg:1997ke}
Forsberg,~N.; Malmqvist,~P.-{\AA}. \emph{Chem. Phys. Lett.} \textbf{1997},
  \emph{274}, 196--204\relax
\mciteBstWouldAddEndPuncttrue
\mciteSetBstMidEndSepPunct{\mcitedefaultmidpunct}
{\mcitedefaultendpunct}{\mcitedefaultseppunct}\relax
\EndOfBibitem
\bibitem[Dyall(1995)]{Dyall:1995ct}
Dyall,~K.~G. \emph{J. Chem. Phys.} \textbf{1995}, \emph{102}, 4909--4918\relax
\mciteBstWouldAddEndPuncttrue
\mciteSetBstMidEndSepPunct{\mcitedefaultmidpunct}
{\mcitedefaultendpunct}{\mcitedefaultseppunct}\relax
\EndOfBibitem
\bibitem[Valdemoro(1992)]{Valdemoro.1992}
Valdemoro,~C. \emph{Phys. Rev. A} \textbf{1992}, \emph{45}, 4462--4467\relax
\mciteBstWouldAddEndPuncttrue
\mciteSetBstMidEndSepPunct{\mcitedefaultmidpunct}
{\mcitedefaultendpunct}{\mcitedefaultseppunct}\relax
\EndOfBibitem
\bibitem[Colmenero and Valdemoro(1993)Colmenero, and Valdemoro]{Colmenero.1993}
Colmenero,~F.; Valdemoro,~C. \emph{Phys. Rev. A} \textbf{1993}, \emph{47},
  979--985\relax
\mciteBstWouldAddEndPuncttrue
\mciteSetBstMidEndSepPunct{\mcitedefaultmidpunct}
{\mcitedefaultendpunct}{\mcitedefaultseppunct}\relax
\EndOfBibitem
\bibitem[Kutzelnigg and Mukherjee(1997)Kutzelnigg, and
  Mukherjee]{kutzelnigg1997normal}
Kutzelnigg,~W.; Mukherjee,~D. \emph{J. Chem. Phys.} \textbf{1997}, \emph{107},
  432--449\relax
\mciteBstWouldAddEndPuncttrue
\mciteSetBstMidEndSepPunct{\mcitedefaultmidpunct}
{\mcitedefaultendpunct}{\mcitedefaultseppunct}\relax
\EndOfBibitem
\bibitem[Mazziotti(1998)]{Mazziotti.1998}
Mazziotti,~D.~A. \emph{Phys. Rev. A} \textbf{1998}, \emph{57}, 4219--4234\relax
\mciteBstWouldAddEndPuncttrue
\mciteSetBstMidEndSepPunct{\mcitedefaultmidpunct}
{\mcitedefaultendpunct}{\mcitedefaultseppunct}\relax
\EndOfBibitem
\bibitem[Mazziotti(2006)]{Mazziotti:2006gf}
Mazziotti,~D.~A. \emph{Phys. Rev. A} \textbf{2006}, \emph{74}, 032501\relax
\mciteBstWouldAddEndPuncttrue
\mciteSetBstMidEndSepPunct{\mcitedefaultmidpunct}
{\mcitedefaultendpunct}{\mcitedefaultseppunct}\relax
\EndOfBibitem
\bibitem[Shamasundar(2009)]{Shamasundar:2009ee}
Shamasundar,~K.~R. \emph{J. Chem. Phys.} \textbf{2009}, \emph{131},
  174109\relax
\mciteBstWouldAddEndPuncttrue
\mciteSetBstMidEndSepPunct{\mcitedefaultmidpunct}
{\mcitedefaultendpunct}{\mcitedefaultseppunct}\relax
\EndOfBibitem
\bibitem[Misiewicz \latin{et~al.}(2020)Misiewicz, Turney, and
  Schaefer]{Misiewicz:2020ia}
Misiewicz,~J.~P.; Turney,~J.~M.; Schaefer,~H.~F. \emph{J. Chem. Theory Comput.}
  \textbf{2020}, \emph{16}, 6150--6164\relax
\mciteBstWouldAddEndPuncttrue
\mciteSetBstMidEndSepPunct{\mcitedefaultmidpunct}
{\mcitedefaultendpunct}{\mcitedefaultseppunct}\relax
\EndOfBibitem
\bibitem[Kurashige \latin{et~al.}(2014)Kurashige, Chalupsky, Lan, and
  Yanai]{Kurashige:2014bq}
Kurashige,~Y.; Chalupsky,~J.; Lan,~T.~N.; Yanai,~T. \emph{J. Chem. Phys.}
  \textbf{2014}, \emph{141}, 174111\relax
\mciteBstWouldAddEndPuncttrue
\mciteSetBstMidEndSepPunct{\mcitedefaultmidpunct}
{\mcitedefaultendpunct}{\mcitedefaultseppunct}\relax
\EndOfBibitem
\bibitem[Phung \latin{et~al.}(2016)Phung, Wouters, and Pierloot]{Phung:2016ke}
Phung,~Q.~M.; Wouters,~S.; Pierloot,~K. \emph{J. Chem. Theory Comput.}
  \textbf{2016}, \emph{12}, 4352--4361\relax
\mciteBstWouldAddEndPuncttrue
\mciteSetBstMidEndSepPunct{\mcitedefaultmidpunct}
{\mcitedefaultendpunct}{\mcitedefaultseppunct}\relax
\EndOfBibitem
\bibitem[Zgid \latin{et~al.}(2009)Zgid, Ghosh, Neuscamman, and
  Chan]{Zgid:2009fu}
Zgid,~D.; Ghosh,~D.; Neuscamman,~E.; Chan,~G. K.-L. \emph{J. Chem. Phys.}
  \textbf{2009}, \emph{130}, 194107\relax
\mciteBstWouldAddEndPuncttrue
\mciteSetBstMidEndSepPunct{\mcitedefaultmidpunct}
{\mcitedefaultendpunct}{\mcitedefaultseppunct}\relax
\EndOfBibitem
\bibitem[Guo \latin{et~al.}(2021)Guo, Sivalingam, and Neese]{Guo2021a}
Guo,~Y.; Sivalingam,~K.; Neese,~F. \emph{J. Chem. Phys.} \textbf{2021},
  \emph{154}, 214111\relax
\mciteBstWouldAddEndPuncttrue
\mciteSetBstMidEndSepPunct{\mcitedefaultmidpunct}
{\mcitedefaultendpunct}{\mcitedefaultseppunct}\relax
\EndOfBibitem
\bibitem[Guo \latin{et~al.}(2021)Guo, Sivalingam, Kollmar, and Neese]{Guo2021b}
Guo,~Y.; Sivalingam,~K.; Kollmar,~C.; Neese,~F. \emph{J. Chem. Phys.}
  \textbf{2021}, \emph{154}, 214113\relax
\mciteBstWouldAddEndPuncttrue
\mciteSetBstMidEndSepPunct{\mcitedefaultmidpunct}
{\mcitedefaultendpunct}{\mcitedefaultseppunct}\relax
\EndOfBibitem
\bibitem[Freitag \latin{et~al.}(2017)Freitag, Knecht, Angeli, and
  Reiher]{Freitag:2017ir}
Freitag,~L.; Knecht,~S.; Angeli,~C.; Reiher,~M. \emph{J. Chem. Theory Comput.}
  \textbf{2017}, \emph{13}, 451--459\relax
\mciteBstWouldAddEndPuncttrue
\mciteSetBstMidEndSepPunct{\mcitedefaultmidpunct}
{\mcitedefaultendpunct}{\mcitedefaultseppunct}\relax
\EndOfBibitem
\bibitem[Sharma and Chan(2014)Sharma, and Chan]{sharma2014communication}
Sharma,~S.; Chan,~G. K.-L. \emph{J. Chem. Phys.} \textbf{2014}, \emph{141},
  111101\relax
\mciteBstWouldAddEndPuncttrue
\mciteSetBstMidEndSepPunct{\mcitedefaultmidpunct}
{\mcitedefaultendpunct}{\mcitedefaultseppunct}\relax
\EndOfBibitem
\bibitem[Sharma \latin{et~al.}(2017)Sharma, Knizia, Guo, and
  Alavi]{sharma2017combining}
Sharma,~S.; Knizia,~G.; Guo,~S.; Alavi,~A. \emph{J. Chem. Theory Comput.}
  \textbf{2017}, \emph{13}, 488--498\relax
\mciteBstWouldAddEndPuncttrue
\mciteSetBstMidEndSepPunct{\mcitedefaultmidpunct}
{\mcitedefaultendpunct}{\mcitedefaultseppunct}\relax
\EndOfBibitem
\bibitem[Sokolov and Chan(2016)Sokolov, and Chan]{sokolov2016time}
Sokolov,~A.~Y.; Chan,~G. K.-L. \emph{J. Chem. Phys.} \textbf{2016}, \emph{144},
  064102\relax
\mciteBstWouldAddEndPuncttrue
\mciteSetBstMidEndSepPunct{\mcitedefaultmidpunct}
{\mcitedefaultendpunct}{\mcitedefaultseppunct}\relax
\EndOfBibitem
\bibitem[Sokolov(2018)]{sokolov2018multi}
Sokolov,~A.~Y. \emph{J. Chem. Phys.} \textbf{2018}, \emph{149}, 204113\relax
\mciteBstWouldAddEndPuncttrue
\mciteSetBstMidEndSepPunct{\mcitedefaultmidpunct}
{\mcitedefaultendpunct}{\mcitedefaultseppunct}\relax
\EndOfBibitem
\bibitem[Chatterjee and Sokolov(2020)Chatterjee, and
  Sokolov]{chatterjee2020extended}
Chatterjee,~K.; Sokolov,~A.~Y. \emph{J. Chem. Theory Comput.} \textbf{2020},
  \emph{16}, 6343--6357\relax
\mciteBstWouldAddEndPuncttrue
\mciteSetBstMidEndSepPunct{\mcitedefaultmidpunct}
{\mcitedefaultendpunct}{\mcitedefaultseppunct}\relax
\EndOfBibitem
\bibitem[Nakano \latin{et~al.}(1998)Nakano, Hirao, and
  Gordon]{nakano1998analytic}
Nakano,~H.; Hirao,~K.; Gordon,~M.~S. \emph{J. Chem. Phys.} \textbf{1998},
  \emph{108}, 5660--5669\relax
\mciteBstWouldAddEndPuncttrue
\mciteSetBstMidEndSepPunct{\mcitedefaultmidpunct}
{\mcitedefaultendpunct}{\mcitedefaultseppunct}\relax
\EndOfBibitem
\bibitem[Nakano(1993)]{Nakano:1993hv}
Nakano,~H. \emph{J. Chem. Phys.} \textbf{1993}, \emph{99}, 7983--7992\relax
\mciteBstWouldAddEndPuncttrue
\mciteSetBstMidEndSepPunct{\mcitedefaultmidpunct}
{\mcitedefaultendpunct}{\mcitedefaultseppunct}\relax
\EndOfBibitem
\bibitem[Nakano \latin{et~al.}(1999)Nakano, Otsuka, and Hirao]{Nakano:2011gl}
Nakano,~H.; Otsuka,~N.; Hirao,~K. \emph{Recent Advances in Multireference
  Methods}; World Scientific, 1999; pp 131--160\relax
\mciteBstWouldAddEndPuncttrue
\mciteSetBstMidEndSepPunct{\mcitedefaultmidpunct}
{\mcitedefaultendpunct}{\mcitedefaultseppunct}\relax
\EndOfBibitem
\bibitem[Park(2021)]{Park2021}
Park,~J.~W. \emph{arXiv} \textbf{2021}, 2106.10817\relax
\mciteBstWouldAddEndPuncttrue
\mciteSetBstMidEndSepPunct{\mcitedefaultmidpunct}
{\mcitedefaultendpunct}{\mcitedefaultseppunct}\relax
\EndOfBibitem
\bibitem[Celani and Werner(2003)Celani, and Werner]{celani2003analytical}
Celani,~P.; Werner,~H.-J. \emph{J. Chem. Phys.} \textbf{2003}, \emph{119},
  5044--5057\relax
\mciteBstWouldAddEndPuncttrue
\mciteSetBstMidEndSepPunct{\mcitedefaultmidpunct}
{\mcitedefaultendpunct}{\mcitedefaultseppunct}\relax
\EndOfBibitem
\bibitem[Dudley \latin{et~al.}(2003)Dudley, Khait, and Hoffmann]{Dudley2003}
Dudley,~T.~J.; Khait,~Y.~G.; Hoffmann,~M.~R. \emph{J. Chem. Phys.}
  \textbf{2003}, \emph{119}, 651--660\relax
\mciteBstWouldAddEndPuncttrue
\mciteSetBstMidEndSepPunct{\mcitedefaultmidpunct}
{\mcitedefaultendpunct}{\mcitedefaultseppunct}\relax
\EndOfBibitem
\bibitem[Theis \latin{et~al.}(2011)Theis, Khait, and Hoffmann]{Theis:2011bd}
Theis,~D.; Khait,~Y.~G.; Hoffmann,~M.~R. \emph{J. Chem. Phys.} \textbf{2011},
  \emph{135}, 044117\relax
\mciteBstWouldAddEndPuncttrue
\mciteSetBstMidEndSepPunct{\mcitedefaultmidpunct}
{\mcitedefaultendpunct}{\mcitedefaultseppunct}\relax
\EndOfBibitem
\bibitem[Khait \latin{et~al.}(2012)Khait, Theis, and Hoffmann]{Khait:2012fi}
Khait,~Y.~G.; Theis,~D.; Hoffmann,~M.~R. \emph{Chem. Phys.} \textbf{2012},
  \emph{401}, 88--94\relax
\mciteBstWouldAddEndPuncttrue
\mciteSetBstMidEndSepPunct{\mcitedefaultmidpunct}
{\mcitedefaultendpunct}{\mcitedefaultseppunct}\relax
\EndOfBibitem
\bibitem[Shiozaki \latin{et~al.}(2011)Shiozaki, Győrffy, Celani, and
  Werner]{shiozaki2011communication}
Shiozaki,~T.; Győrffy,~W.; Celani,~P.; Werner,~H.-J. \emph{J. Chem. Phys.}
  \textbf{2011}, \emph{135}, 081106\relax
\mciteBstWouldAddEndPuncttrue
\mciteSetBstMidEndSepPunct{\mcitedefaultmidpunct}
{\mcitedefaultendpunct}{\mcitedefaultseppunct}\relax
\EndOfBibitem
\bibitem[Gy{\H o}rffy \latin{et~al.}(2013)Gy{\H o}rffy, Shiozaki, Knizia, and
  Werner]{Gyorffy:2013kf}
Gy{\H o}rffy,~W.; Shiozaki,~T.; Knizia,~G.; Werner,~H.-J. \emph{J. Chem. Phys.}
  \textbf{2013}, \emph{138}, 104104\relax
\mciteBstWouldAddEndPuncttrue
\mciteSetBstMidEndSepPunct{\mcitedefaultmidpunct}
{\mcitedefaultendpunct}{\mcitedefaultseppunct}\relax
\EndOfBibitem
\bibitem[MacLeod and Shiozaki(2015)MacLeod, and
  Shiozaki]{macleod2015communication}
MacLeod,~M.~K.; Shiozaki,~T. \emph{J. Chem. Phys.} \textbf{2015}, \emph{142},
  051103\relax
\mciteBstWouldAddEndPuncttrue
\mciteSetBstMidEndSepPunct{\mcitedefaultmidpunct}
{\mcitedefaultendpunct}{\mcitedefaultseppunct}\relax
\EndOfBibitem
\bibitem[Vlaisavljevich and Shiozaki(2016)Vlaisavljevich, and
  Shiozaki]{Vlaisavljevich:2016kv}
Vlaisavljevich,~B.; Shiozaki,~T. \emph{J. Chem. Theory Comput.} \textbf{2016},
  \emph{12}, 3781--3787\relax
\mciteBstWouldAddEndPuncttrue
\mciteSetBstMidEndSepPunct{\mcitedefaultmidpunct}
{\mcitedefaultendpunct}{\mcitedefaultseppunct}\relax
\EndOfBibitem
\bibitem[Park and Shiozaki(2017)Park, and Shiozaki]{Park:2017je}
Park,~J.~W.; Shiozaki,~T. \emph{J. Chem. Theory Comput.} \textbf{2017},
  \emph{13}, 2561--2570\relax
\mciteBstWouldAddEndPuncttrue
\mciteSetBstMidEndSepPunct{\mcitedefaultmidpunct}
{\mcitedefaultendpunct}{\mcitedefaultseppunct}\relax
\EndOfBibitem
\bibitem[Park \latin{et~al.}(2019)Park, Al-Saadon, Strand, and
  Shiozaki]{Park:2019fh}
Park,~J.~W.; Al-Saadon,~R.; Strand,~N.~E.; Shiozaki,~T. \emph{J. Chem. Theory
  Comput.} \textbf{2019}, \emph{15}, 4088--4098\relax
\mciteBstWouldAddEndPuncttrue
\mciteSetBstMidEndSepPunct{\mcitedefaultmidpunct}
{\mcitedefaultendpunct}{\mcitedefaultseppunct}\relax
\EndOfBibitem
\bibitem[Shiozaki(2018)]{Shiozaki:2018fm}
Shiozaki,~T. \emph{Wiley Interdiscip. Rev. Comput. Mol. Sci.} \textbf{2018},
  \emph{8}, e1331\relax
\mciteBstWouldAddEndPuncttrue
\mciteSetBstMidEndSepPunct{\mcitedefaultmidpunct}
{\mcitedefaultendpunct}{\mcitedefaultseppunct}\relax
\EndOfBibitem
\bibitem[Song \latin{et~al.}(2021)Song, Neaton, and Mart{\'\i}nez]{Song:2021hp}
Song,~C.; Neaton,~J.~B.; Mart{\'\i}nez,~T.~J. \emph{J. Chem. Phys.}
  \textbf{2021}, \emph{154}, 014103\relax
\mciteBstWouldAddEndPuncttrue
\mciteSetBstMidEndSepPunct{\mcitedefaultmidpunct}
{\mcitedefaultendpunct}{\mcitedefaultseppunct}\relax
\EndOfBibitem
\bibitem[Song \latin{et~al.}(2021)Song, Mart{\'{i}}nez, and Neaton]{Song2021}
Song,~C.; Mart{\'{i}}nez,~T.~J.; Neaton,~J.~B. \emph{J. Chem. Phys.}
  \textbf{2021}, \emph{155}, 024108\relax
\mciteBstWouldAddEndPuncttrue
\mciteSetBstMidEndSepPunct{\mcitedefaultmidpunct}
{\mcitedefaultendpunct}{\mcitedefaultseppunct}\relax
\EndOfBibitem
\bibitem[Park(2019)]{Park2019}
Park,~J.~W. \emph{J. Chem. Theory Comput.} \textbf{2019}, \emph{15},
  5417--5425\relax
\mciteBstWouldAddEndPuncttrue
\mciteSetBstMidEndSepPunct{\mcitedefaultmidpunct}
{\mcitedefaultendpunct}{\mcitedefaultseppunct}\relax
\EndOfBibitem
\bibitem[Park(2020)]{Park2020}
Park,~J.~W. \emph{J. Chem. Theory Comput.} \textbf{2020}, \emph{16},
  326--339\relax
\mciteBstWouldAddEndPuncttrue
\mciteSetBstMidEndSepPunct{\mcitedefaultmidpunct}
{\mcitedefaultendpunct}{\mcitedefaultseppunct}\relax
\EndOfBibitem
\bibitem[Nishimoto(2019)]{Nishimoto:2019hd}
Nishimoto,~Y. \emph{J. Chem. Phys.} \textbf{2019}, \emph{151}, 114103\relax
\mciteBstWouldAddEndPuncttrue
\mciteSetBstMidEndSepPunct{\mcitedefaultmidpunct}
{\mcitedefaultendpunct}{\mcitedefaultseppunct}\relax
\EndOfBibitem
\bibitem[Evangelista(2014)]{evangelista2014driven}
Evangelista,~F.~A. \emph{J. Chem. Phys.} \textbf{2014}, \emph{141},
  054109\relax
\mciteBstWouldAddEndPuncttrue
\mciteSetBstMidEndSepPunct{\mcitedefaultmidpunct}
{\mcitedefaultendpunct}{\mcitedefaultseppunct}\relax
\EndOfBibitem
\bibitem[Li and Evangelista(2019)Li, and Evangelista]{Li:2019fu}
Li,~C.; Evangelista,~F.~A. \emph{Annu. Rev. Phys. Chem.} \textbf{2019},
  \emph{70}, 245--273\relax
\mciteBstWouldAddEndPuncttrue
\mciteSetBstMidEndSepPunct{\mcitedefaultmidpunct}
{\mcitedefaultendpunct}{\mcitedefaultseppunct}\relax
\EndOfBibitem
\bibitem[Li and Evangelista(2015)Li, and Evangelista]{li2015multireference}
Li,~C.; Evangelista,~F.~A. \emph{J. Chem. Theory Comput.} \textbf{2015},
  \emph{11}, 2097--2108\relax
\mciteBstWouldAddEndPuncttrue
\mciteSetBstMidEndSepPunct{\mcitedefaultmidpunct}
{\mcitedefaultendpunct}{\mcitedefaultseppunct}\relax
\EndOfBibitem
\bibitem[Li and Evangelista(2016)Li, and Evangelista]{Li:2016hb}
Li,~C.; Evangelista,~F.~A. \emph{J. Chem. Phys.} \textbf{2016}, \emph{144},
  164114\relax
\mciteBstWouldAddEndPuncttrue
\mciteSetBstMidEndSepPunct{\mcitedefaultmidpunct}
{\mcitedefaultendpunct}{\mcitedefaultseppunct}\relax
\EndOfBibitem
\bibitem[Li and Evangelista(2017)Li, and Evangelista]{li2017driven}
Li,~C.; Evangelista,~F.~A. \emph{J. Chem. Phys.} \textbf{2017}, \emph{146},
  124132\relax
\mciteBstWouldAddEndPuncttrue
\mciteSetBstMidEndSepPunct{\mcitedefaultmidpunct}
{\mcitedefaultendpunct}{\mcitedefaultseppunct}\relax
\EndOfBibitem
\bibitem[Zhang \latin{et~al.}(2019)Zhang, Li, and Evangelista]{Zhang:2019ec}
Zhang,~T.; Li,~C.; Evangelista,~F.~A. \emph{J. Chem. Theory Comput.}
  \textbf{2019}, \emph{15}, 4399--4414\relax
\mciteBstWouldAddEndPuncttrue
\mciteSetBstMidEndSepPunct{\mcitedefaultmidpunct}
{\mcitedefaultendpunct}{\mcitedefaultseppunct}\relax
\EndOfBibitem
\bibitem[Li and Evangelista(2021)Li, and Evangelista]{Li:2021a}
Li,~C.; Evangelista,~F.~A. \emph{J. Chem. Phys.} \textbf{2021}, \emph{155},
  114111\relax
\mciteBstWouldAddEndPuncttrue
\mciteSetBstMidEndSepPunct{\mcitedefaultmidpunct}
{\mcitedefaultendpunct}{\mcitedefaultseppunct}\relax
\EndOfBibitem
\bibitem[Hannon \latin{et~al.}(2016)Hannon, Li, and
  Evangelista]{hannon2016integral}
Hannon,~K.~P.; Li,~C.; Evangelista,~F.~A. \emph{J. Chem. Phys.} \textbf{2016},
  \emph{144}, 204111\relax
\mciteBstWouldAddEndPuncttrue
\mciteSetBstMidEndSepPunct{\mcitedefaultmidpunct}
{\mcitedefaultendpunct}{\mcitedefaultseppunct}\relax
\EndOfBibitem
\bibitem[Schriber \latin{et~al.}(2018)Schriber, Hannon, Li, and
  Evangelista]{Schriber:2018hw}
Schriber,~J.~B.; Hannon,~K.~P.; Li,~C.; Evangelista,~F.~A. \emph{J. Chem.
  Theory Comput.} \textbf{2018}, \emph{14}, 6295--6305\relax
\mciteBstWouldAddEndPuncttrue
\mciteSetBstMidEndSepPunct{\mcitedefaultmidpunct}
{\mcitedefaultendpunct}{\mcitedefaultseppunct}\relax
\EndOfBibitem
\bibitem[Khokhlov and Belov(2021)Khokhlov, and Belov]{Khokhlov2021}
Khokhlov,~D.; Belov,~A. \emph{J. Chem. Theory Comput.} \textbf{2021},
  \emph{17}, 4301--4315\relax
\mciteBstWouldAddEndPuncttrue
\mciteSetBstMidEndSepPunct{\mcitedefaultmidpunct}
{\mcitedefaultendpunct}{\mcitedefaultseppunct}\relax
\EndOfBibitem
\bibitem[Helgaker and J{\o}rgensen(1988)Helgaker, and
  J{\o}rgensen]{helgaker1988analytical}
Helgaker,~T.; J{\o}rgensen,~P. \emph{Adv. Quantum Chem.} \textbf{1988},
  \emph{19}, 183--245\relax
\mciteBstWouldAddEndPuncttrue
\mciteSetBstMidEndSepPunct{\mcitedefaultmidpunct}
{\mcitedefaultendpunct}{\mcitedefaultseppunct}\relax
\EndOfBibitem
\bibitem[Helgaker(1998)]{Helgaker:2002hs}
Helgaker,~T. In \emph{Encyclopedia of Computational Chemistry}; Schleyer,~P.
  v.~R., Allinger,~N.~L., Clark,~T., Gasteiger,~J., Kollman,~P.~A.,
  Schaefer,~H.~F., Schreiner,~P.~R., Eds.; John Wiley {\&} Sons, Ltd:
  Chichester, UK, 1998; pp 1157--1169\relax
\mciteBstWouldAddEndPuncttrue
\mciteSetBstMidEndSepPunct{\mcitedefaultmidpunct}
{\mcitedefaultendpunct}{\mcitedefaultseppunct}\relax
\EndOfBibitem
\bibitem[Wang \latin{et~al.}(2019)Wang, Li, and Evangelista]{wang2019analytic}
Wang,~S.; Li,~C.; Evangelista,~F.~A. \emph{J. Chem. Phys.} \textbf{2019},
  \emph{151}, 044118\relax
\mciteBstWouldAddEndPuncttrue
\mciteSetBstMidEndSepPunct{\mcitedefaultmidpunct}
{\mcitedefaultendpunct}{\mcitedefaultseppunct}\relax
\EndOfBibitem
\bibitem[Handy and Schaefer(1984)Handy, and Schaefer]{handy1984evaluation}
Handy,~N.~C.; Schaefer,~H.~F. \emph{J. Chem. Phys.} \textbf{1984}, \emph{81},
  5031--5033\relax
\mciteBstWouldAddEndPuncttrue
\mciteSetBstMidEndSepPunct{\mcitedefaultmidpunct}
{\mcitedefaultendpunct}{\mcitedefaultseppunct}\relax
\EndOfBibitem
\bibitem[Yamaguchi \latin{et~al.}(1994)Yamaguchi, Osamura, Goddard, and
  Schaefer]{Yamaguchi1994book}
Yamaguchi,~Y.; Osamura,~Y.; Goddard,~J.~D.; Schaefer,~H.~F. \emph{A New
  Dimension to Quantum Chemistry: Analytic Derivative Methods in Ab Initio
  Molecular Electronic Structure Theory}; Oxford University Press, 1994\relax
\mciteBstWouldAddEndPuncttrue
\mciteSetBstMidEndSepPunct{\mcitedefaultmidpunct}
{\mcitedefaultendpunct}{\mcitedefaultseppunct}\relax
\EndOfBibitem
\bibitem[Osamura \latin{et~al.}(1982)Osamura, Yamaguchi, and
  Schaefer]{Osamura.1982}
Osamura,~Y.; Yamaguchi,~Y.; Schaefer,~H.~F. \emph{J. Chem. Phys.}
  \textbf{1982}, \emph{77}, 383--390\relax
\mciteBstWouldAddEndPuncttrue
\mciteSetBstMidEndSepPunct{\mcitedefaultmidpunct}
{\mcitedefaultendpunct}{\mcitedefaultseppunct}\relax
\EndOfBibitem
\bibitem[Rice and Amos(1985)Rice, and Amos]{Rice:1985gg}
Rice,~J.~E.; Amos,~R.~D. \emph{Chem. Phys. Lett.} \textbf{1985}, \emph{122},
  585--590\relax
\mciteBstWouldAddEndPuncttrue
\mciteSetBstMidEndSepPunct{\mcitedefaultmidpunct}
{\mcitedefaultendpunct}{\mcitedefaultseppunct}\relax
\EndOfBibitem
\bibitem[Levchenko \latin{et~al.}(2005)Levchenko, Wang, and
  Krylov]{levchenko2005analytic}
Levchenko,~S.~V.; Wang,~T.; Krylov,~A.~I. \emph{J. Chem. Phys.} \textbf{2005},
  \emph{122}, 224106\relax
\mciteBstWouldAddEndPuncttrue
\mciteSetBstMidEndSepPunct{\mcitedefaultmidpunct}
{\mcitedefaultendpunct}{\mcitedefaultseppunct}\relax
\EndOfBibitem
\bibitem[Roos \latin{et~al.}(1980)Roos, Taylor, and Sigbahn]{roos1980complete}
Roos,~B.~O.; Taylor,~P.~R.; Sigbahn,~P.~E. \emph{Chem. Phys.} \textbf{1980},
  \emph{48}, 157--173\relax
\mciteBstWouldAddEndPuncttrue
\mciteSetBstMidEndSepPunct{\mcitedefaultmidpunct}
{\mcitedefaultendpunct}{\mcitedefaultseppunct}\relax
\EndOfBibitem
\bibitem[Werner and Knowles(1990)Werner, and Knowles]{werner1990comparison}
Werner,~H.-J.; Knowles,~P.~J. \emph{Theor. Chim. Acta.} \textbf{1990},
  \emph{78}, 175--187\relax
\mciteBstWouldAddEndPuncttrue
\mciteSetBstMidEndSepPunct{\mcitedefaultmidpunct}
{\mcitedefaultendpunct}{\mcitedefaultseppunct}\relax
\EndOfBibitem
\bibitem[FOR(2021)]{FORTE2021}
Forte, a suite of quantum chemistry methods for strongly correlated electrons.
  For current version see \url{https://github.com/evangelistalab/forte},
  2021\relax
\mciteBstWouldAddEndPuncttrue
\mciteSetBstMidEndSepPunct{\mcitedefaultmidpunct}
{\mcitedefaultendpunct}{\mcitedefaultseppunct}\relax
\EndOfBibitem
\bibitem[Smith \latin{et~al.}(2020)Smith, Burns, Simmonett, Parrish, Schieber,
  Galvelis, Kraus, Kruse, {Di Remigio}, Alenaizan, James, Lehtola, Misiewicz,
  Scheurer, Shaw, Schriber, Xie, Glick, Sirianni, O'Brien, Waldrop, Kumar,
  Hohenstein, Pritchard, Brooks, Schaefer, Sokolov, Patkowski, DePrince,
  Bozkaya, King, Evangelista, Turney, Crawford, and Sherrill]{smith2020psi4}
Smith,~D. G.~A.; Burns,~L.~A.; Simmonett,~A.~C.; Parrish,~R.~M.;
  Schieber,~M.~C.; Galvelis,~R.; Kraus,~P.; Kruse,~H.; {Di Remigio},~R.;
  Alenaizan,~A.; James,~A.~M.; Lehtola,~S.; Misiewicz,~J.~P.; Scheurer,~M.;
  Shaw,~R.~A.; Schriber,~J.~B.; Xie,~Y.; Glick,~Z.~L.; Sirianni,~D.~A.;
  O'Brien,~J.~S.; Waldrop,~J.~M.; Kumar,~A.; Hohenstein,~E.~G.;
  Pritchard,~B.~P.; Brooks,~B.~R.; Schaefer,~H.~F.; Sokolov,~A.~Y.;
  Patkowski,~K.; DePrince,~A.~E.; Bozkaya,~U.; King,~R.~A.; Evangelista,~F.~A.;
  Turney,~J.~M.; Crawford,~T.~D.; Sherrill,~C.~D. \emph{J. Chem. Phys.}
  \textbf{2020}, \emph{152}, 184108\relax
\mciteBstWouldAddEndPuncttrue
\mciteSetBstMidEndSepPunct{\mcitedefaultmidpunct}
{\mcitedefaultendpunct}{\mcitedefaultseppunct}\relax
\EndOfBibitem
\bibitem[Dunning(1989)]{dunning1989a}
Dunning,~T.~H. \emph{J. Chem. Phys.} \textbf{1989}, \emph{90}, 1007--1023\relax
\mciteBstWouldAddEndPuncttrue
\mciteSetBstMidEndSepPunct{\mcitedefaultmidpunct}
{\mcitedefaultendpunct}{\mcitedefaultseppunct}\relax
\EndOfBibitem
\bibitem[Woon and Dunning(1995)Woon, and Dunning]{woon1995a}
Woon,~D.~E.; Dunning,~T.~H. \emph{J. Chem. Phys.} \textbf{1995}, \emph{103},
  4572--4585\relax
\mciteBstWouldAddEndPuncttrue
\mciteSetBstMidEndSepPunct{\mcitedefaultmidpunct}
{\mcitedefaultendpunct}{\mcitedefaultseppunct}\relax
\EndOfBibitem
\bibitem[Mahapatra \latin{et~al.}(1998)Mahapatra, Datta, and
  Mukherjee]{mahapatra1998state}
Mahapatra,~U.~S.; Datta,~B.; Mukherjee,~D. \emph{Mol. Phys.} \textbf{1998},
  \emph{94}, 157--171\relax
\mciteBstWouldAddEndPuncttrue
\mciteSetBstMidEndSepPunct{\mcitedefaultmidpunct}
{\mcitedefaultendpunct}{\mcitedefaultseppunct}\relax
\EndOfBibitem
\bibitem[Evangelista \latin{et~al.}(2010)Evangelista, Prochnow, Gauss, and
  Schaefer]{evangelista2010perturbative}
Evangelista,~F.~A.; Prochnow,~E.; Gauss,~J.; Schaefer,~H.~F. \emph{J. Chem.
  Phys.} \textbf{2010}, \emph{132}, 074107\relax
\mciteBstWouldAddEndPuncttrue
\mciteSetBstMidEndSepPunct{\mcitedefaultmidpunct}
{\mcitedefaultendpunct}{\mcitedefaultseppunct}\relax
\EndOfBibitem
\bibitem[Werner \latin{et~al.}(2015)Werner, Knowles, Knizia, Manby,
  {Sch\"{u}tz}, Celani, Gy{\H o}rffy, Kats, Korona, Lindh, Mitrushenkov,
  Rauhut, Shamasundar, Adler, Amos, Bernhardsson, Berning, Cooper, Deegan,
  Dobbyn, Eckert, Goll, Hampel, Hesselmann, Hetzer, Hrenar, Jansen, K\"oppl,
  Liu, Lloyd, Mata, May, McNicholas, Meyer, Mura, Nicklass, O'Neill, Palmieri,
  Peng, Pfl\"uger, Pitzer, Reiher, Shiozaki, Stoll, Stone, Tarroni,
  Thorsteinsson, and Wang]{MOLPRO2015}
Werner,~H.-J.; Knowles,~P.~J.; Knizia,~G.; Manby,~F.~R.; {Sch\"{u}tz},~M.;
  Celani,~P.; Gy{\H o}rffy,~W.; Kats,~D.; Korona,~T.; Lindh,~R.;
  Mitrushenkov,~A.; Rauhut,~G.; Shamasundar,~K.~R.; Adler,~T.~B.; Amos,~R.~D.;
  Bernhardsson,~A.; Berning,~A.; Cooper,~D.~L.; Deegan,~M. J.~O.;
  Dobbyn,~A.~J.; Eckert,~F.; Goll,~E.; Hampel,~C.; Hesselmann,~A.; Hetzer,~G.;
  Hrenar,~T.; Jansen,~G.; K\"oppl,~C.; Liu,~Y.; Lloyd,~A.~W.; Mata,~R.~A.;
  May,~A.~J.; McNicholas,~S.~J.; Meyer,~W.; Mura,~M.~E.; Nicklass,~A.;
  O'Neill,~D.~P.; Palmieri,~P.; Peng,~D.; Pfl\"uger,~K.; Pitzer,~R.;
  Reiher,~M.; Shiozaki,~T.; Stoll,~H.; Stone,~A.~J.; Tarroni,~R.;
  Thorsteinsson,~T.; Wang,~M. MOLPRO, version 2015.1, a package of $ab$
  $initio$ programs. 2015; see http://www.molpro.net\relax
\mciteBstWouldAddEndPuncttrue
\mciteSetBstMidEndSepPunct{\mcitedefaultmidpunct}
{\mcitedefaultendpunct}{\mcitedefaultseppunct}\relax
\EndOfBibitem
\bibitem[Li and Evangelista(2018)Li, and Evangelista]{Li:2018kl}
Li,~C.; Evangelista,~F.~A. \emph{J. Chem. Phys.} \textbf{2018}, \emph{148},
  124106\relax
\mciteBstWouldAddEndPuncttrue
\mciteSetBstMidEndSepPunct{\mcitedefaultmidpunct}
{\mcitedefaultendpunct}{\mcitedefaultseppunct}\relax
\EndOfBibitem
\end{mcitethebibliography}

\end{document}